\documentclass[journal,draftcls,onecolumn,12pt,twoside]{IEEEtran}
\usepackage{amssymb}
\usepackage{amsmath}
\usepackage[version=4]{mhchem}
\usepackage{graphicx}
\usepackage{multirow}
\usepackage{wrapfig}
\usepackage{color}
\usepackage{relsize}
\usepackage{siunitx}
\usepackage{colortbl}
\usepackage{bm}
\usepackage{soul}
\usepackage[noadjust]{cite}
\usepackage[flushleft]{threeparttable}
\usepackage{subfigure}
\usepackage{enumerate}
\usepackage{paralist}
\usepackage{dsfont}
\usepackage[normalem]{ulem}
\usepackage{mathtools}
\usepackage{nameref,hyperref}
\usepackage{amssymb}
\usepackage{amsbsy}
\usepackage{amsmath}
\usepackage{graphicx}
\usepackage{multirow}
\usepackage{wrapfig}
\usepackage{color}
\usepackage{colortbl}
\usepackage{soul}
\usepackage{subfigure}
\usepackage{enumerate}
\usepackage{paralist}
\usepackage{rotating}
\usepackage{tikz}
\usepackage[normalem]{ulem}
\usepackage{etaremune}
\usepackage{mathtools}
\usepackage{stfloats}
\usepackage{booktabs} % For professional looking tables
\usepackage{multirow}
\usepackage[flushleft]{threeparttable}
\usepackage{siunitx} % use this package module for SI units
\usepackage[version=4]{mhchem}
\usepackage{relsize}
\usepackage{bm}
\usepackage{dsfont}
\usepackage{tabularx,booktabs,caption}
\usepackage[flushleft]{threeparttable}

\def\mathclap#1{\text{\hbox to 0pt{\hss$\mathsurround=0pt#1$\hss}}}
\makeatletter
\newcommand{\vastt}{\bBigg@{3}}
\newcommand{\vast}{\bBigg@{4}}
\newcommand{\Vast}{\bBigg@{5}}
\makeatother

\begin{document}
\title{Graphene-based Nanoscale Molecular Communication Receiver: Fabrication and Microfluidic Analysis} 
\author{Murat Kuscu, Hamideh Ramezani, Ergin Dinc, Shahab Akhavan, Ozgur B. Akan
	\thanks{Murat Kuscu and Hamideh Ramezani are with the Internet of Everything (IoE) Group and Cambridge Graphene Centre (CGC), Department of Engineering, University of Cambridge, Cambridge, CB3 0FA, UK (e-mail: \{mk959, hr404\}@cam.ac.uk).}
	\thanks{Ergin Dinc and Ozgur B. Akan are with the Internet of Everything (IoE) Group, Department of Engineering, University of Cambridge, Cambridge, CB3 0FA, UK (e-mail: \{ed502, oba21\}@cam.ac.uk).}
	\thanks{Shahab Akhavan is with the Cambridge Graphene Centre, (CGC) Department of Engineering, University of Cambridge, Cambridge, CB3 0FA, UK (e-mail: sa766@cam.ac.uk).}	
%	\thanks{Ozgur B. Akan is also with the Department of Electrical and Electronics Engineering, Koc University, Istanbul, 34450, Turkey  (email: akan@ku.edu.tr).}
	\thanks{This work was supported by the ERC project MINERVA (ERC-2013-CoG \#616922).}}% <-this % stops a space

\maketitle

\begin{abstract}
Bio-inspired molecular communications (MC), where molecules are used to transfer information, is the most promising technique to realise the Internet of Nano Things (IoNT), thanks to its inherent biocompatibility, energy-efficiency, and reliability in physiologically-relevant environments. Despite a substantial body of theoretical work concerning MC, the lack of practical micro/nanoscale MC devices and MC testbeds has led researchers to make overly simplifying assumptions about the implications of the channel conditions and the physical architectures of the practical transceivers in developing theoretical models and devising communication methods for MC. On the other hand, MC imposes unique challenges resulting from the highly complex, nonlinear, time-varying channel properties that cannot be always tackled by conventional information and communication tools and technologies (ICT). As a result, the reliability of the existing MC methods, which are mostly adopted from electromagnetic communications and not validated with practical testbeds, is highly questionable. As the first step to remove this discrepancy, in this study, we report on the fabrication of a nanoscale MC receiver based on graphene field-effect transistor biosensors. We perform its ICT characterisation in a custom-designed microfluidic MC system with the information encoded into the concentration of single-stranded DNA molecules. This experimental platform is the first practical implementation of a micro/nanoscale MC system with nanoscale MC receivers, and can serve as a testbed for developing realistic MC methods and IoNT applications.
\end{abstract}

\IEEEpeerreviewmaketitle
%\begin{IEEEkeywords}
%Molecular communications, receiver, graphene, biosensor, DNA, microfluidics
%\end{IEEEkeywords}

\section{Introduction}

Nanotechnology is enabling us to devise ever-smaller devices to interact with the universe at molecular resolution. Though small enough to penetrate into cells, these devices individually are of limited capability. To unleash the full potential of nanotechnology, communication among nanomachines is a must. This would enable more complex applications, e.g., continuous health monitoring with intrabody nanosensor networks, theranostic applications with distributed sensor and actuator networks, and industrial nanosensor applications. Internet of Nano Things (IoNT), which defines these networks of nanomachines, such as nanobiosensors and engineered bacteria, integrated with the Internet infrastructure, has seen tremendous interest recently \cite{akyildiz2010internet, akyildiz2015internet,  dinc2019internet}. IoNT is set to transform the way we connect with and understand the world at the bottom. However, conventional electromagnetic (EM) communication techniques, proved impractical at nanoscale due to the antenna size and power limitations, and large propagation losses. On the other hand, Nature itself already provides a robust way of nanocommunication, i.e., Molecular Communications (MC), which is the common communication modality among living cells, ranging from our own neurons to bacteria \cite{akan2017fundamentals}. Using the same language with living cells by encoding, transmitting and receiving information with molecules provides an energy efficient and reliable nanocommunication, even at harsh biological environments, where, we expect, the most impactful medical applications of IoNT would be implemented.

Although there is tremendous interest in this field accompanied by a large body of theoretical work to develop models and devise communication methods for MC \cite{pierobon2011diffusion, pierobon2011noise, kuran2011modulation, kuscu2016modeling, bilgin2018dna, kuscu2018maximum, kuscu2019channel}, researchers rely on overly simplifying assumptions about the physical constraints of the transceivers and the channel conditions due to the lack of a testbed for MC at micro/nanoscale, where these models and methods can be validated. On the other hand, MC brings about unique challenges resulting from its highly complex, nonlinear, time-varying channel properties due to the discrete nature of information carriers (molecules), substantial channel memory and peculiarities of molecular interactions at nanoscale, that cannot be always tackled by conventional ICT tools \cite{akan2017fundamentals, kuscu2019transmitter,jamali2019channel}. This leaves a huge question mark over the reliability of the existing MC methods, which are mostly adopted from conventional EM communications and not validated with practical testbeds. 

Few studies in MC literature have focused on \emph{macroscale} implementation of MC systems taking into account the physical limitations of a receiver, although the utilized receivers are made of off-the-shelf macroscale components. In \cite{farsad2013tabletop}, the isopropyl alcohol (IPA) is used as information carrier, and commercially available metal oxide semiconductor alcohol sensors are used as MC receiver. This study provides a testbed for MC with macroscale dimensions, which is later utilised in \cite{kim2015universal} to estimate its combined channel and receiver model. This testbed is extended to a molecular multiple-input multiple-output (MIMO) system in \cite{koo2016molecular} to improve the achievable data rate. In \cite{farsad2017novel}, the information is encoded in pH level of the transmitted fluid, and a pH probe sensor is used as the MC receiver. On the grounds that the use of acids and bases for information transmission can adversely affect the other processes in the application environment, such as in the human body, magnetic nanoparticles (MNs) are employed as information-carrying molecules in microfluidic channels in \cite{unterweger2018experimental}. In that study, a bulky susceptometer is used to detect the concentration of MNs and decode the transmitted messages. In addition, the performance of MN-based MC, where an external magnetic field is employed to attract the MNs to a passive receiver, is analysed in \cite{wicke2018magnetic}. However, the focus of these works is on macroscale MC using off-the-shelf sensors as receiver. Therefore, these studies do not contribute to the development of a design and optimisation framework for practical nanoscale MC receivers that can actually be integrated into micro/nanoscale devices.

As the first step to overcome this challenge, in this work, we report on the first implementation of a nanoscale MC receiver based on graphene field-effect transistor-based DNA biosensors (graphene bioFETs), and its ICT performance tests in a custom-designed microfluidic MC system. The main objective of this work is to provide an experimental testbed at physically relevant dimensions for nanonetworks, which can be used to reveal and study the effects of intricate biochemical and physical processes on the MC performance, and develop practical and realistic communication methods, including new MC detection techniques. 

Graphene, with its exceptional electrical, chemical and mechanical properties, such as high carrier mobility at room temperature, one atomic layer thickness and two dimensional geometry exposing all its atoms to the sensing environment, provides very high sensitivity towards biochemical molecules especially in a bioFET configuration \cite{ferrari2015science, huang2010nanoelectronic, zhan2014graphene}. Owing to these properties, graphene has been extensively studied for selective sensing of a wide range of biomolecules ranging from carbohydrates \cite{kwak2012flexible} to proteins \cite{ohno2009electrolyte} and oligonucleotides \cite{xu2017real,campos2019attomolar}. Meeting the fundamental requirements of an MC receiver, such as the capability of label-free and reversible detection and high sensitivity \cite{kuscu2016physical}, graphene bioFET stands as an ideal candidate for the implementation of the MC receiver. Flexibility and nanoscale 2d geometry of graphene are particularly favourable for the integration of graphene-based MC receiver into functional nanoscale devices.

Functionalisation of graphene with biomolecular probes can provide the selectivity against target analytes, required for avoiding biochemical interference for MC applications in physiologically relevant environments. In this work, graphene is functionalised with single-stranded DNA (ssDNA) probes (pDNAs) which undergo reversible hybridisation reaction with the complementary target DNAs (tDNAs). The reason for selecting DNA as the recognition element is that DNA can be easily customized with different base sequences of different lengths, and can be designed to bind not only complementary DNAs but also peptides, proteins, carbohydrates and small molecules \cite{green2015interactions}. As a result, the implemented device can serve as a model system to provide insight into a broad range of MC systems relying on detection of molecular messages through affinity-based ligand-receptor interactions \cite{pierobon2011noise,kuscu2019transmitter}. Moreover, integration of the fabricated MC receiver into a pressure-regulated microfluidic testbed provides control over the fluid flow rate, and enables flexibility and practicality in testing different channel geometries, which can mimic the most promising application environments of the MC inside human body, e.g., circulatory system \cite{malak2012molecular, chahibi2014molecular}.

In the remainder of this paper, we elaborate on the fabrication process of single layer graphene (SLG) bioFET-based MC receiver and its integration into a microfluidic testbed. The electrical characterisation of the device is performed at each step of functionalisation. Sensing response characteristics are revealed to determine the affinity between the complementary tDNA-pDNA pair. Selectivity of the device against complementary tDNAs is examined through real-time sensing response to non-complementary target DNAs (ntDNAs). Following the fabrication and sensitivity/selectivity analysis, we provide an MC detection performance analysis based on the transmission of pseudo-random binary data encoded into the concentration of tDNAs. The time-varying response of the MC receiver is fitted by a previously developed microfluidic MC model. This analysis provides important insights particularly into the infamous ISI problem of MC resulting from the slow kinetics of ligand-receptor binding reactions. Similar to the existing approaches in the MC literature \cite{deng2015modeling,li2015low}, a concentration difference-based detection method is utilised to overcome the ISI effects by obviating the need for channel state information (CSI).

\section{Fabrication of MC Receiver}

MC receiver is fabricated in three consecutive steps. First, a graphene field-effect transistor (GFET) with chemical vapour deposition (CVD)-grown SLG is fabricated on Si/SiO$_2$ substrate through optical lithography techniques. Then, a polydimethylsiloxane (PDMS)-based microfluidic channel is produced to encapsulate the GFET for bio-functionalisation, and real-time microfluidic sensing and communication experiments. Finally, for selectivity of the MC receiver against information-carrying target DNAs, bio-functionalisation of the GFET channels with probe DNA molecules is performed inside the microfluidic channel connected to a pressure-regulated microfluidic setup.

MC receiver is fabricated with a CVD-grown SLG polycrystalline domain on an n-type Si/SiO$_2$ substrate (525 $\mu$m with 90 nm thermal oxide layer, obtained from Mi-Net Technology Ltd). The CVD-grown SLG on Cu with PMMA coating (60 nm, 495K, A2) is obtained from Graphenea Inc. 

\subsubsection{Wet Transfer of CVD Graphene on Si/SiO$_2$ Substrate} The Cu layer of the PMMA/SLG/Cu stack should be removed before transfer onto Si/SiO$_2$ substrate. The Cu side is partially covered with a graphitic film, which may result in poor Cu etching performance. This backside graphene on Cu is etched away with O$_2$ plasma at 3 W for 30 seconds in a low-power Reactive Ion Etcher (RIE) in NanoEtch (Moorfield Nanotechnology Ltd). After this step, Cu is etched by placing the PMMA/SLG/Cu stack on the surface of a solution of ammonium persulphate (APS) (1.8 g of APS in 150 ml DI water (18.2 M$\Omega$$\cdot$cm)).
\begin{figure}[!b]
	\centering
	\includegraphics[width=15cm]{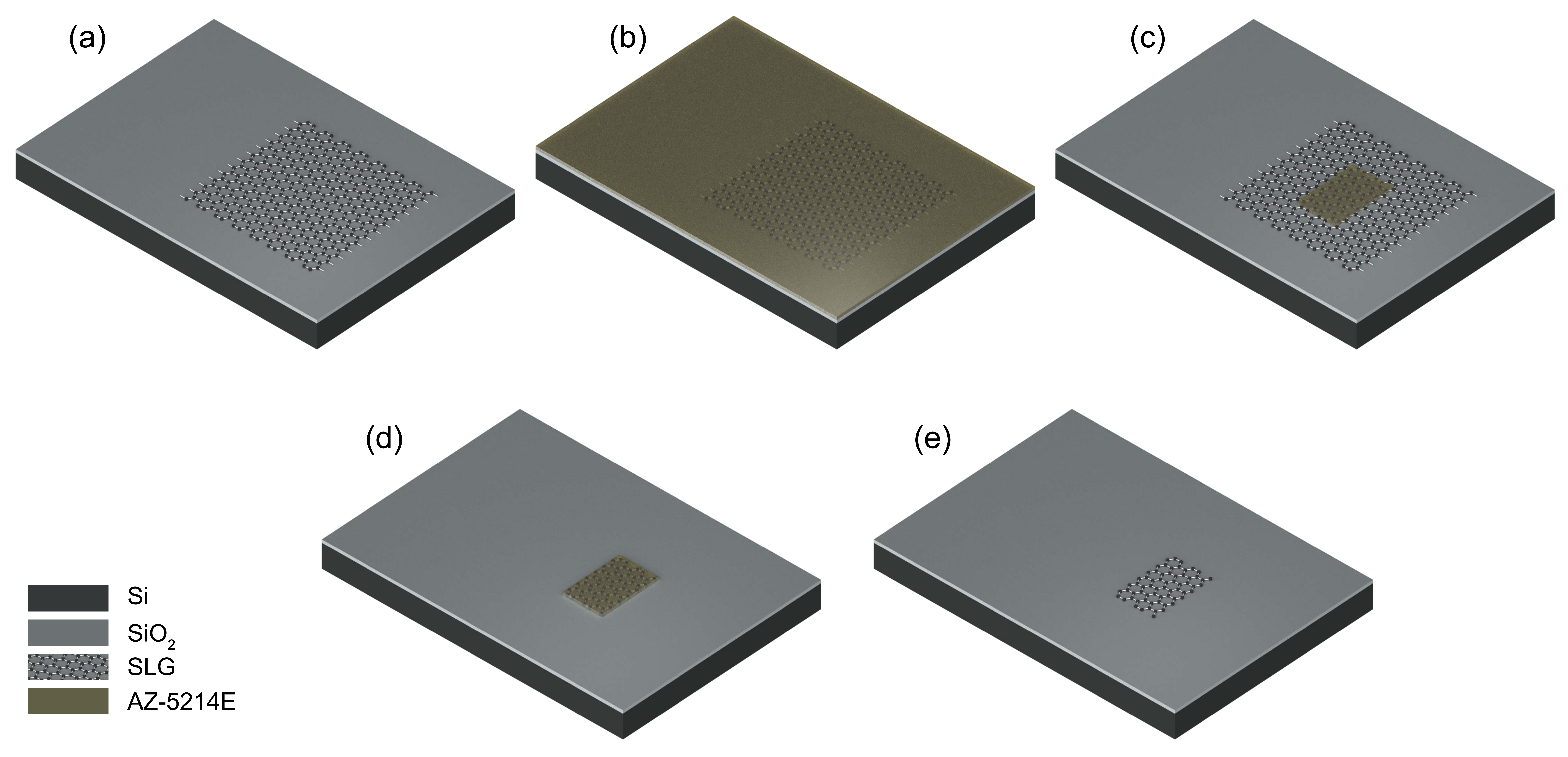}
	\caption{Process flow for the patterning of SLG channels. (a) SLG transferred on Si/SiO$_2$ substrate. (b) Coating of sample with photoresist layer. (c) SLG channel pattern defined by optical lithography. (d) SLG channel pattern after RIE etching. (e) Removal of residual photoresist layer from SLG surface, and the resulting SLG channel on the substrate.}
	\label{fig:fab_gr_pattern}
\end{figure}

The Si/SiO$_2$ substrate is cleaned before the transfer by means of sonication for 10 minutes in acetone followed by immersion in isopropyl alcohol (IPA) for 5 minutes and drying with nitrogen (N$_2$). Once the Cu is entirely dissolved, the floating PMMA/SLG stack is transferred onto the surface of DI water in a beaker by a glass slide to dilute the APS residuals, and then, the stack is fished onto the Si/SiO$_2$ substrate. The resulting sample (PMMA/SLG/SiO$_2$/Si) is left vertically to dry overnight, and then annealed over a hot plate at 150 $^\circ$C for 2 hours. The sample is then transferred into a beaker with acetone for PMMA removal for 2 hours, and then immersed in IPA for 5 minutes and dried with N$_2$, leaving only the SLG film on the Si/SiO$_2$ substrate (Fig. \ref{fig:fab_gr_pattern}(a)).

\begin{figure}[!t]
	\centering
	\includegraphics[width=13cm]{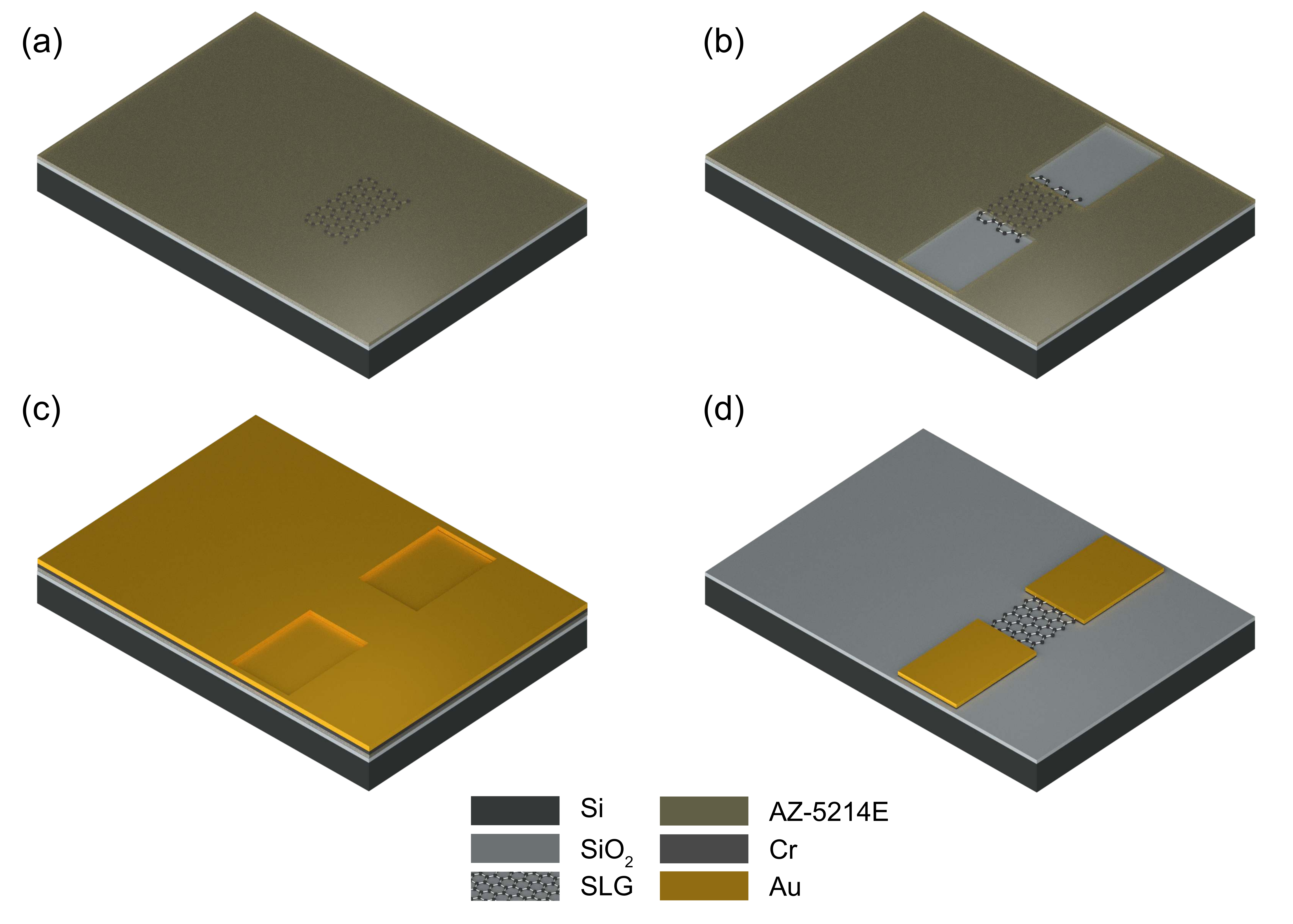}
	\caption{Process flow for the deposition of contacts. (a) Coating of sample with photoresist layer. (b) Contact pattern defined by optical lithography. (c) Deposition of Cr and Au metal films through thermal evaporation. (d) Patterned contacts after lift-off process.}
	\label{fig:fab_contact}
\end{figure}

\subsubsection{Patterning of SLG Channels} MC receiver is designed to contain 7 GFETs having isolated source and drain contacts but being exposed to a common electrolyte gate. The individual SLG channels are patterned via optical lithography with a laser writer according to the design shown in Fig. \ref{fig:litho_layers} in Appendix \ref{AppendixB}. Prior to all laser writing processes in this work, the sample is spin-coated with a photoresist (AZ-5214E from Microchemicals GmbH) at 4000 rpm for 60 seconds (Fig. \ref{fig:fab_gr_pattern}(b)), and baked at 110 $^\circ$C for 50 seconds on a hot plate. The photoresist layer is exposed by direct laser writing (wavelength-405 nm 169 mJ/cm$^2$) via laser writer (LW-405B+ from Microtech Srl), and the pattern is successively developed in diluted developer solution (1:4, AZ-351B/DI Water) for 35-45 seconds followed by brief immersion of the sample in DI-water for 2 seconds and drying with N$_2$ (Fig. \ref{fig:fab_gr_pattern}(c)). The patterning of the individual SLG channels is completed with RIE removing the undesired areas of SLG film, which are not covered with the photoresist layer, via O$_2$ plasma at 3 W for 60 seconds (Fig. \ref{fig:fab_gr_pattern}(d)).  Finally, to remove the photoresist layer, the sample is successively immersed in acetone and IPA for 20 minutes and 5 minutes, respectively, and dried with N$_2$, leaving the patterned SLG film on the substrate (Fig. \ref{fig:fab_gr_pattern}(e)). 

\subsubsection{Deposition of Contacts} In the following step, metal contact areas (source and drain) are defined on the sample through another optical lithography process (Figs. \ref{fig:fab_contact}(a)-(b)). Depositions of 5 nm Cr and 50 nm Au are performed successively over the sample covered with the patterned photoresist layer by thermal evaporation at 10$^{-6}$ mbar (using MiniLab 060 from Moorfield Nanotechnology Ltd) (Fig. \ref{fig:fab_contact}(c)). Once the metals are deposited uniformly, the sample is dipped in acetone for 2 hours for lift-off process, during which the metals over the photoresist layer are removed leaving only the patterned contacts (Fig. \ref{fig:fab_contact}(d)).

\begin{figure}[!t]
	\centering
	\includegraphics[width=15cm]{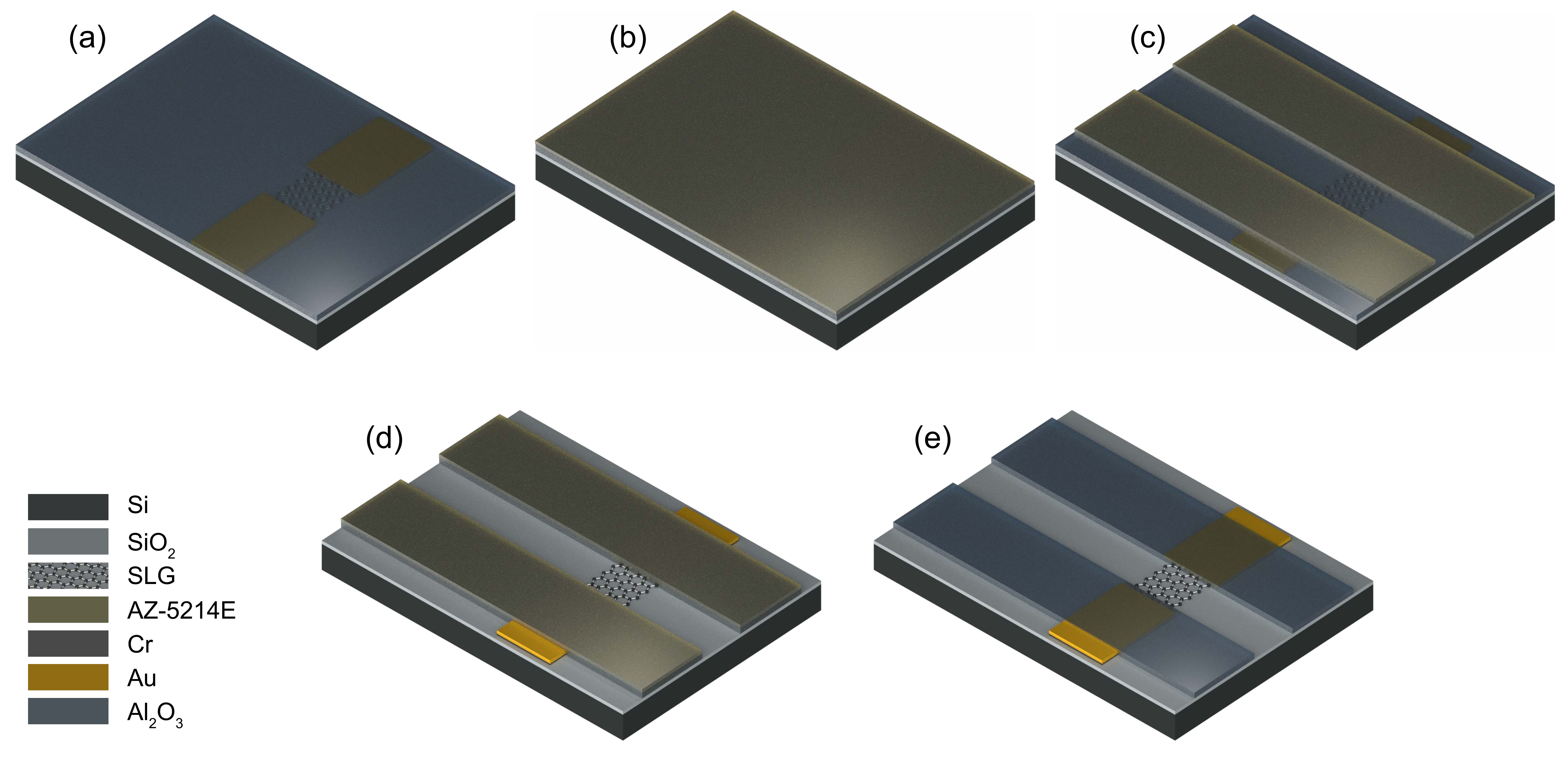}
	\caption{Process flow for the deposition of insulator. (a) Uniform Al$_2$O$_3$ film over the sample after ALD. (b) Coating of sample with photoresist layer. (c) Insulator pattern defined by optical lithography. (d) Exposed SLG channel and contacts after wet etching of Al$_2$O$_3$. (e) Patterned Al$_2$O$_3$ film after removal of excess photoresist. }
	\label{fig:fab_ALD}
\end{figure}

\begin{figure}[t!]
	\centering
	\includegraphics[width=10cm]{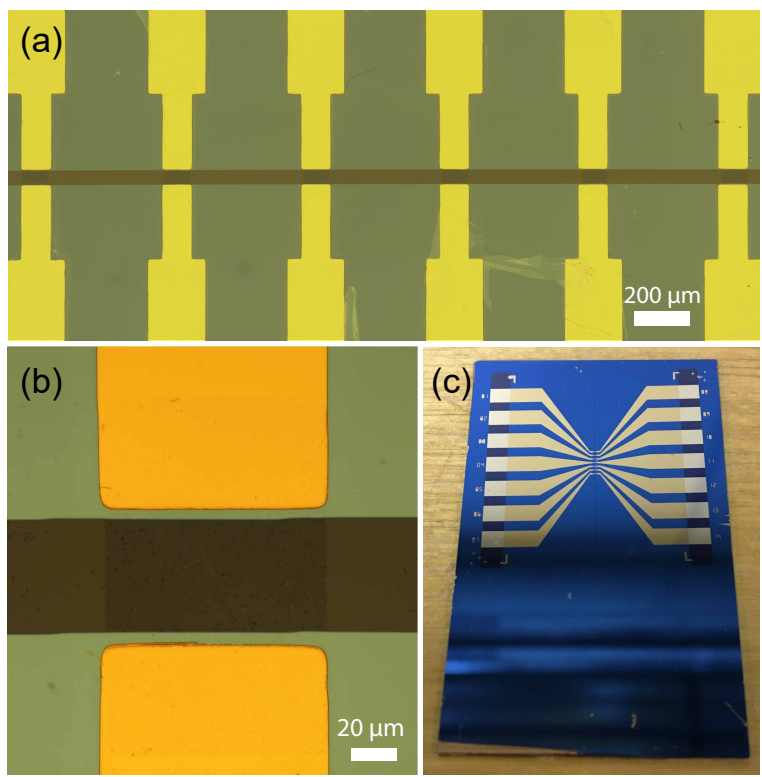}
	\caption{(a) Optical micrograph of the fabricated GFET channels after Al$_2$O$_3$ etching process (only six of the seven channels are visible). (b) A closer look into one of the GFET channels. (c) Overall view of the fabricated 7-channel GFET before the bonding of microfluidic PDMS layer. }\label{fig:grapheneFET}
\end{figure}

\subsubsection{Deposition of Insulator} Finally, the drain and source contact areas, which might be exposed to the electrolyte during microfluidic experiments, are insulated to prevent any parasitic current between metal contacts through the electrolyte. For this, a thin layer of Al$_2$O$_3$ (20 nm) is uniformly deposited over the sample through atomic layer deposition (ALD) in TFS200 (manufactured by Beneq) (Fig. \ref{fig:fab_ALD}(a)). This step is followed by another optical lithography process, which defines the windows over Al$_2$O$_3$ to expose only the SLG channels to the electrolyte, and to expose the source and drain contact pads, which remain outside of the microfluidic channel for electrical measurements (Figs. \ref{fig:fab_ALD}(b)-(c)). The exposed areas of Al$_2$O$_3$ are then wet-etched in Phosphoric Acid (85\% wt. in H$_2$O obtained from Sigma-Aldrich) at 60 $^\circ$C for 1 minute (Fig. \ref{fig:fab_ALD}(d)). For removing any excess photoresist, the sample is dipped in acetone for 20 minutes and IPA for 5 minutes followed by drying with N$_2$, leaving the patterned Al$_2$O$_3$ film on top (Fig. \ref{fig:fab_ALD}(e)). The optical images of the fabricated GFET are shown in Fig. \ref{fig:grapheneFET}.

\subsection{Fabrication of Microfluidic Channels and Device Integration}
\begin{figure}[!t]
	\centering
	\includegraphics[width=15cm]{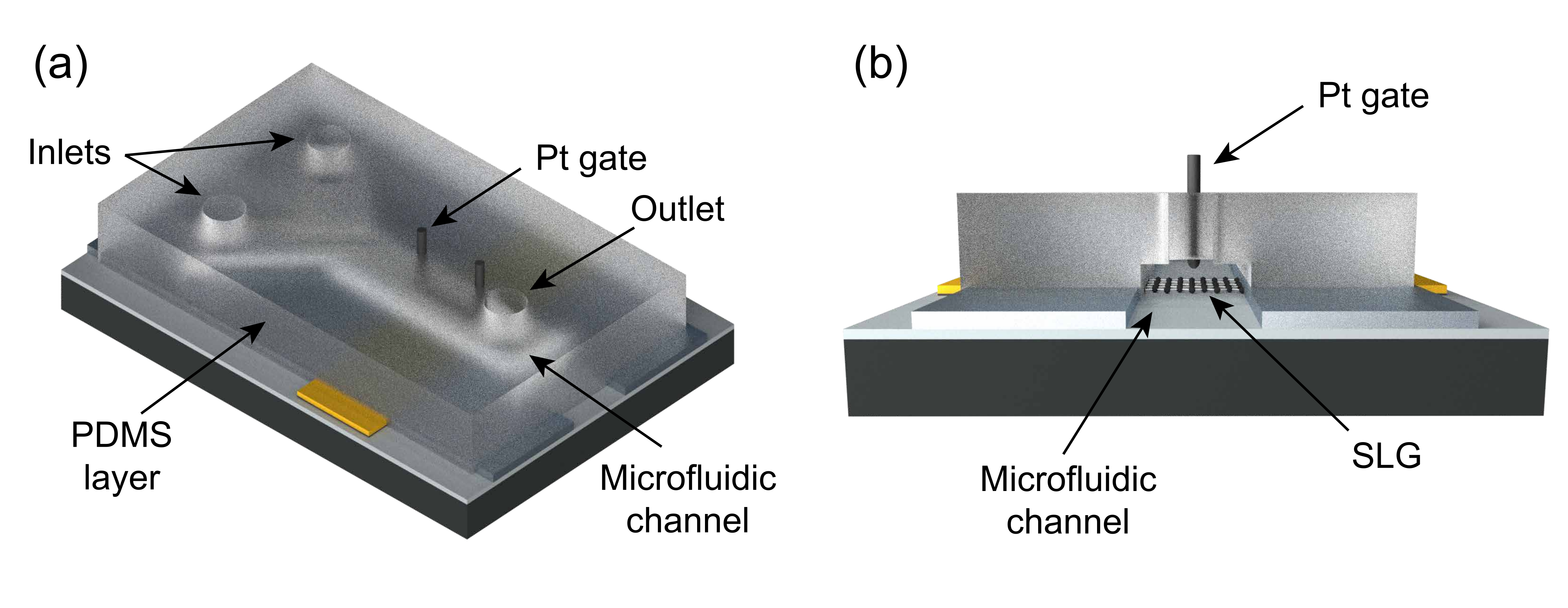}
	\caption{(a) Microfluidic PDMS layer bonded to the GFET surface after the inlets and outlet are defined, and the Pt gate electrode is placed on top. (b) Cross-sectional view of the MC receiver after PDMS layer bonding.} 
	\label{fig:fab_PDMS}
\end{figure}

Fabricated GFET is encapsulated with a PDMS microfluidic channel, as demonstrated in Fig. \ref{fig:fab_PDMS}. To this end, a 3d-printed mould is designed to define the geometry of the rectangular fluidic channel within the PDMS layer (see Fig. \ref{fig:PDMS_mould}(a) in Appendix \ref{AppendixB}). The microfluidic channel has a width of $4\times10^3$ $\mu$m and a height of $1.5\times10^3$ $\mu$m. At one end, the channel bifurcates for connection with the two channel inlets, which are designated for connection to the fluid reservoirs containing the buffer and information-carrying tDNA solutions during the communication experiments.

PDMS prepolymer is prepared using a 10:1 mixture of PDMS base monomer (Sylgard 184 Silicone Elastomer) and PDMS curing agent (obtained from Dow Corning Corporation). Air bubbles inside the PDMS are removed by degassing in a desiccator for 1 hour. The degassed mixture is poured onto the  3d-printed mould and left for curing overnight at room temperature. The cured PDMS is then carefully peeled off from its mould (see Fig. \ref{fig:PDMS_mould}(b) in Appendix \ref{AppendixB}). The inlet and outlet holes are punched through the PDMS layer for microfluidic connections by a biopsy punch (1.25 mm radius). A platinum (Pt) wire having a diameter of 0.5 mm acting as the common solution gate is then mounted to the top of the PDMS channel right above the SLG channels (Fig. \ref{fig:fab_PDMS}, and Fig. \ref{fig:PDMS_mould}(c) in Appendix \ref{AppendixB}). The length of the Pt wire inside the channel is set to 1 cm. 

\begin{figure}[!t]
	\centering
	\includegraphics[width=14.5cm]{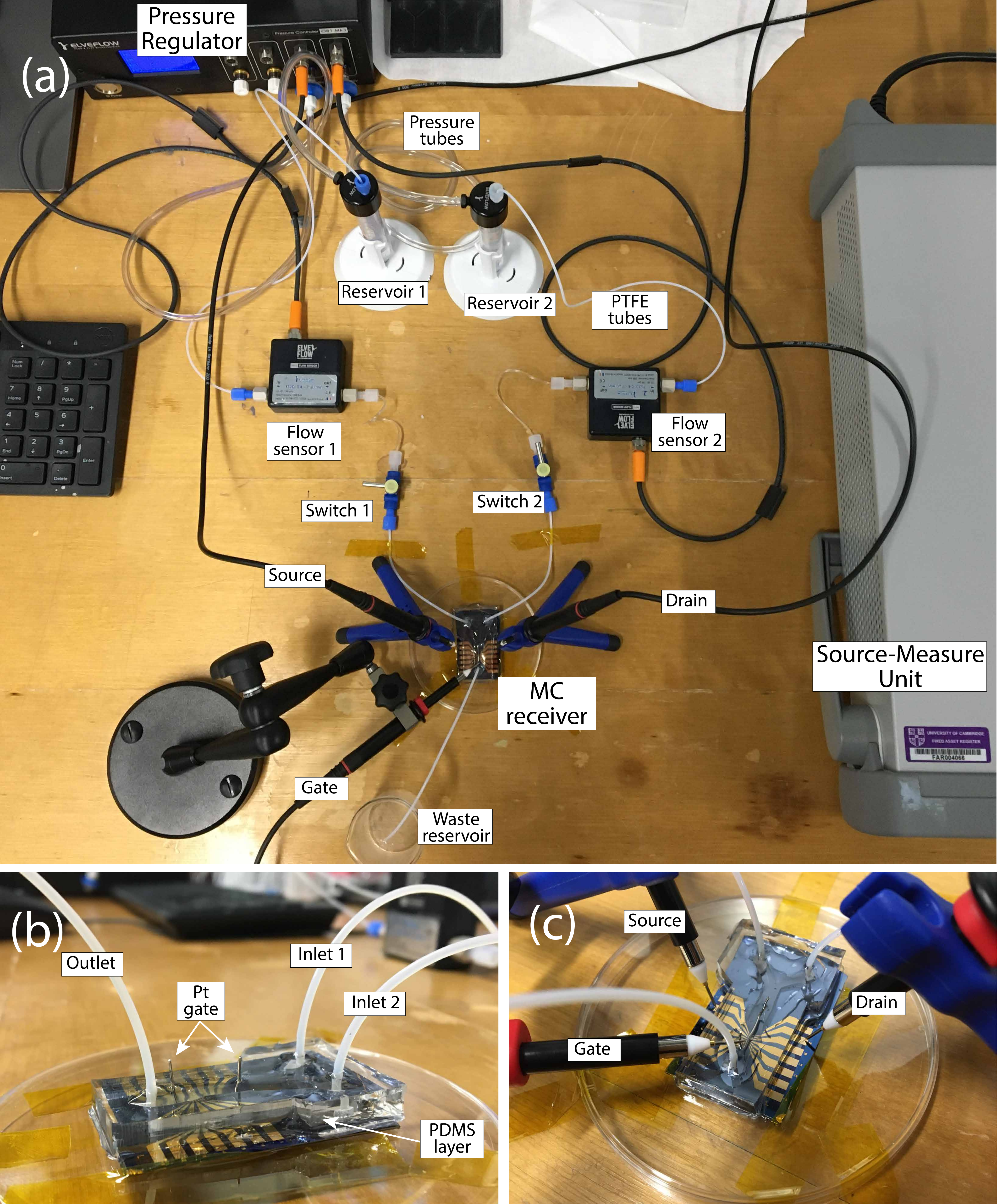}
	\caption{(a) Microfluidic measurement setup consisting of a 4-channel pressure regulator, a high-precision SMU, electrical probes and microfluidic accessories. (b) A closer look into the fabricated graphene-based MC receiver connected to the microfluidic setup. (c) Probe connections for electrical tests of the device.  }	
	\label{fig:microfluidic_setup2}
\end{figure}

In the next step, the patterned PDMS with the Pt solution gate is bonded to the surface of the MC receiver, ensuring that the graphene channels are well-aligned with the microfluidic channel and not placed under the PDMS walls. The most common method for bonding PDMS on SiO$_2$ and glass substrates is based on the O$_2$ plasma activation of the PDMS surface and the target substrate. This requires the surfaces of both the PDMS and the target substrate to be smooth. In our case, however, the exposed graphene channels on the target SiO$_2$ substrate prevents the application of the O$_2$ plasma, as this would cause the removal of graphene channels through plasma etching. Moreover, the plasma activation of only the PDMS surface is not sufficient because curing in the 3d-printed moulds made of Polylactic acid (PLA) results in PDMS layers with a rough surface (see Fig. \ref{fig:PDMS_mould}(c) in Appendix \ref{AppendixB}) rendering O$_2$ plasma activation ineffective in bonding. Therefore, we apply an alternative method, which was first introduced in \cite{chueh2007leakage} for bonding porous membranes into PDMS devices. In this method, a thin layer of PDMS prepolymer, which is in liquid form, is coated on the bonding surface of the cured PDMS layer. Then, the PDMS is carefully placed on the sample, which is cleaned off any dust with N$_2$ prior to bonding. After placement, it takes approximately 1 minute for the PDMS prepolymer to spread uniformly and cover the entire area between the PDMS and substrate except for the empty area defining the microfluidic channel. Once a uniform PDMS prepolymer layer is observed, the temperature of the hot plate is increased to 150 $^\circ$C, and the prepolymer, serving as mortar, is quickly cured, resulting in a strong bonding. This method has consistently yielded leakage-free PDMS-substrate bonding during the fabrication process.

After bonding process, the inlet and outlet tubes are placed on the predefined inlet/outlet holes, as shown in Fig. \ref{fig:microfluidic_setup2}(b). Here, Teflon PTFE tubing (1/16'' OD x 1/32'' ID, obtained from Darwin Microfluidics) is preferred because of its higher chemical stability compared to Tygon tubing, which reacts with DMF used in the functionalisation process. The placement of the inlet and outlet tubes is followed by the application of PDMS prepolymer around the connection points of inlet, outlet and Pt gate over the cured PDMS layer for the complete sealing of the device.

The dimensions of the microfluidic channel together with the fluid flow rate and fluid properties determine the Reynolds number, which is a dimensionless variable indicating the fluid flow regime in the channel \cite{bicen2013system}. Reynolds number is the ratio of the inertial forces to the viscous forces, and can be given by
\begin{align}
	\mathrm{Re} = \frac{\rho u D_H}{\mu}, \label{Reynolds}
\end{align}
where $\rho$ is the fluid density, $u$ is the linear flow velocity of the fluid, $\mu$ is the viscosity of the fluid, and $D_H$ is the hydraulic diameter, which can be obtained for rectangular channels as follows
\begin{align}
	D_H = \frac{4 A_{ch}}{P}. \label{Reynolds}
\end{align}
Here $A_{ch} = w_{ch} \times h_{ch}$ is the cross-sectional area of the channel, and $P = 2(w_{ch}+h_{ch})$ is the cross-sectional channel perimeter. In the sensing and communication experiments of this work, water-based solutions are flowed at a constant volumetric flow rate $u_V = 80$ $\mu$l/min. The linear flow velocity can then be obtained as $u = u_V/A_{ch} = 220$ $\mu$m/s. By using $\rho \approx 1000$ kg$\cdot$m$^{-3}$ and $\mu \approx 0.001002$ Pa$\cdot$s for water, we can obtain the Reynolds number for the microfluidic MC system as $\mathrm{Re} = 0.4839$, indicating a strong laminar flow regime, where viscous forces overcome the inertial forces resulting in non-crossing, parallel streamlines \cite{bicen2013system}. 
 
\begin{figure*}[!b]
	\centering
	\includegraphics[width=13cm]{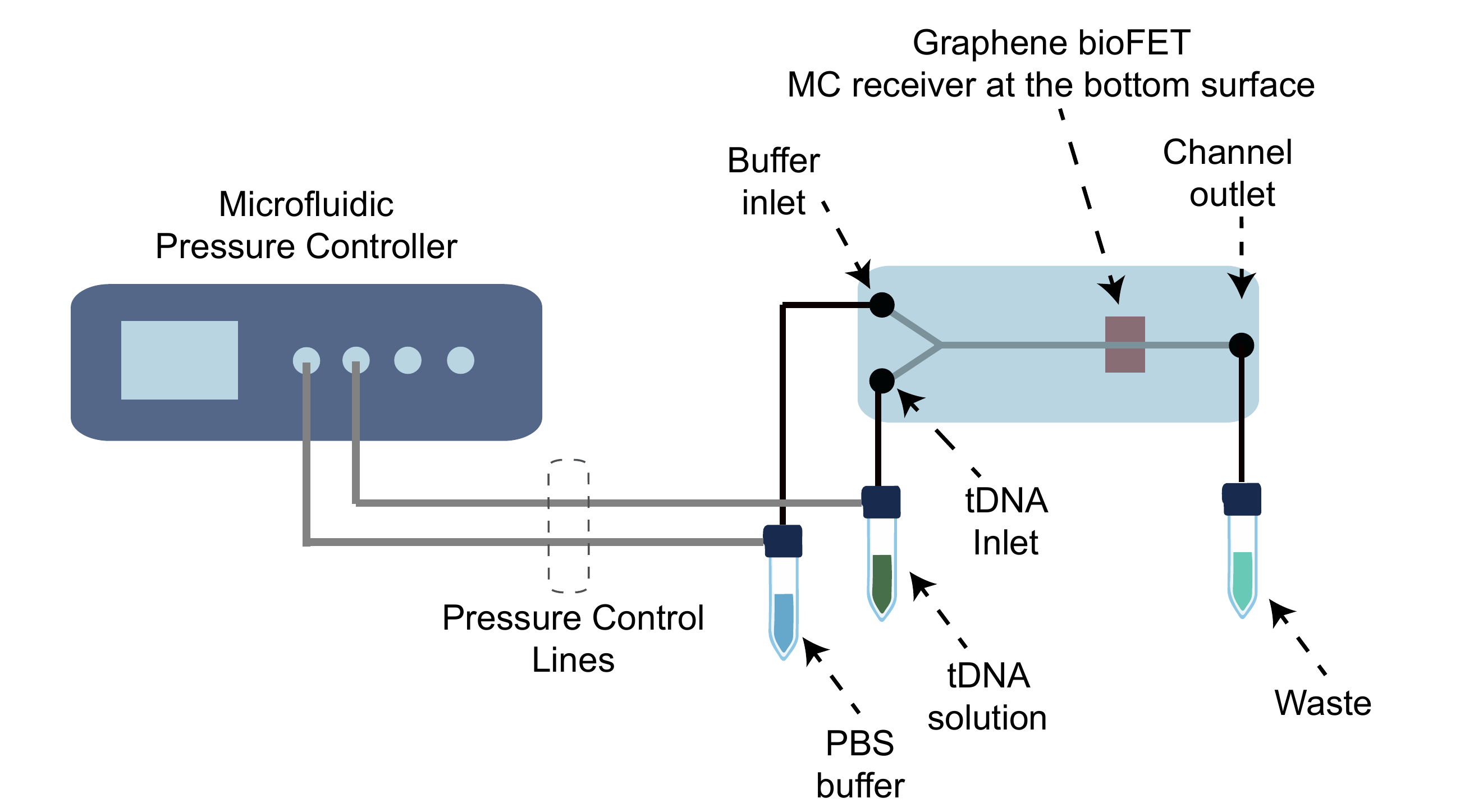}
	\caption{Conceptual drawing of the microfluidic measurement setup with the practical implementation shown in Fig. \ref{fig:microfluidic_setup2}} 
	\label{fig:microfluidic_setup}
\end{figure*}

\subsubsection{Microfluidic Setup}
For functionalisation and electrical characterisation in the next steps, the device was connected to a microfluidic test setup, as shown in Fig. \ref{fig:microfluidic_setup2}. The setup consists of a pressure regulator (OB1 MK3 - Microfluidic flow control system, obtained from Elveflow) with four pressure outlets, two of which are connected to fluid reservoirs through the pressure inlets. The fluid outlets of the fluid reservoirs are connected to the device through PTFE tubing. 

Throughout the bio-functionalisation, sensing and communication experiments, microfluidic flow sensors are partly utilised for feedback-controlled modulation of the inlet pressure, and mechanical flow switches are used in cases where immediate stop/start of the microfluidic flow is required. 

\subsection{Functionalisation of GFET}

Due to its one atomic thickness and 2d structure, the electronic properties of the pristine SLG is highly sensitive to the biochemical environment in the vicinity of its surface. Therefore, it suffers from low-level selectivity. On the other hand, in order to suppress the interference from other biochemical processes in physiologically relevant applications of the MC receiver, selectivity against information-carrying molecules is a must.  Selectivity of the graphene can be realised through bio-functionalisation with recognition elements such as DNA and antibodies. As ssDNAs are preferred as target information-carrying molecules, i.e., tDNAs, in this work, the fabricated GFET is functionalised with probe DNAs (pDNAs), which are complementary to tDNAs. 

\begin{figure}[!t]
	\centering
	\includegraphics[width=16cm]{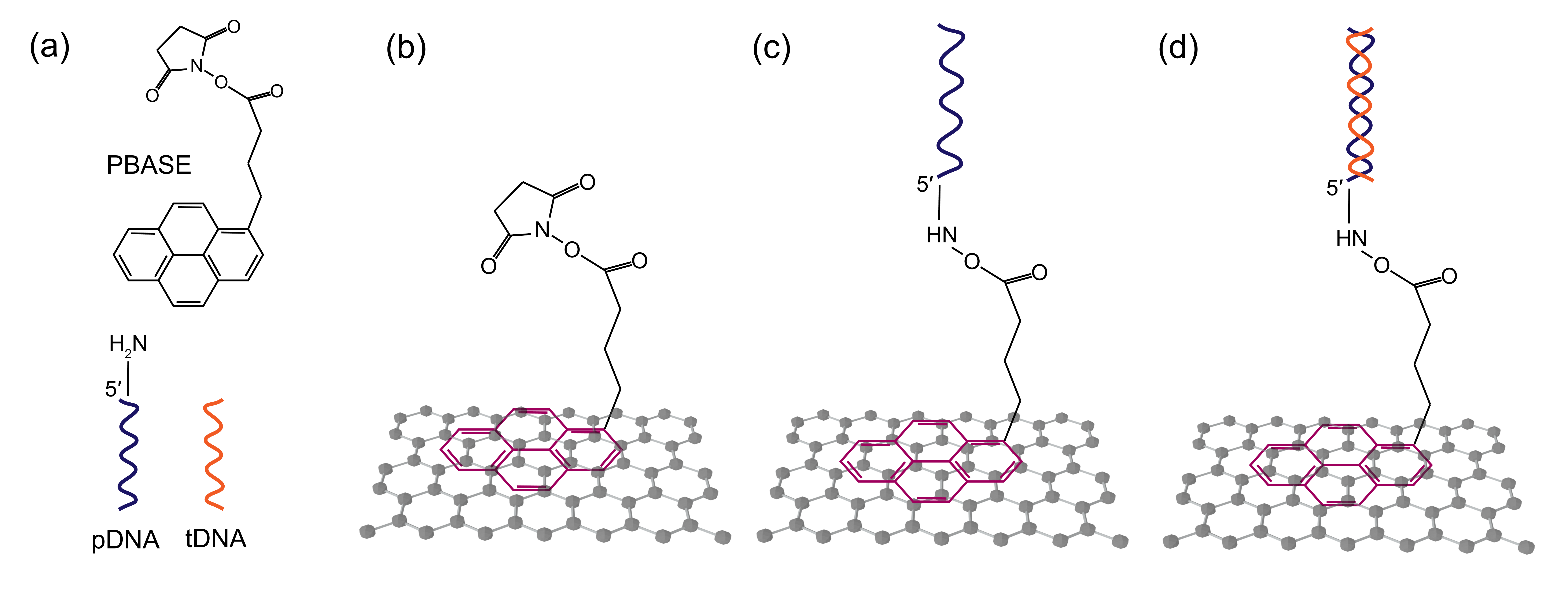}
	\caption{(a) Molecular structure of PBASE, and conceptual drawing of probe DNA (pDNA) and complementary target DNA (tDNA). (b) Noncovalent binding of PBASE to graphene via $\pi-\pi$ interaction. (c) Immobilisation of pDNA via conjugation reaction with the succinimide group of PBASE. (d) pDNA-tDNA hybridisation.  } 
	\label{fig:functionalization}
\end{figure}

For increasing the strength of the probe DNA immobilisation, and reducing the effect of nonspecific binding, the pristine SLG channels are first functionalised with 1-Pyrenebutyric acid N-hydroxysuccinimide ester (PBASE, obtained from Cambridge Bioscience Ltd), which has been widely utilised in the literature as linker molecules between graphene surface and DNA molecules \cite{xu2017real, campos2019attomolar, yue2017electricity, hwang2016highly}. PBASE is an aromatic molecule having an aromatic pyrenyl group and an amine-reactive succinimide group (Fig. \ref{fig:functionalization}(a)). PBASE exhibits a strong affinity towards SLG as its aromatic pyrenyl group interacts with the basal plane of graphene through  $\pi$-$\pi$ interactions resulting in a strong noncovalent binding (Fig. \ref{fig:functionalization}(b)). The noncovalent attachment of the PBASE does not alter the inherent electronic structure and physical properties of the graphene \cite{wu2017doping}. For the functionalisation of the SLG with PBASE molecules, 10 mM solution of PBASE in N,N-Dimethylformamide (DMF, anhydrous, 99.8\%, obtained from Sigma-Aldrich) is prepared in a glass bottle, and sonicated for 30 seconds for mixing. The prepared PBASE/DMF solution is flowed through the microfluidic channel until the entire channel is filled with the solution. Then the flow is stopped, and the SLG channels are exposed to steady PBASE/DMF solution for 2 hours. After functionalisation with PBASE, unbound PBASE molecules are removed from the channel with pure DMF, followed by rinsing with phosphate buffered saline (PBS,  pH 7.4). 

The next step is the immobilisation of 18-mer 5'-amine-modified probe DNAs, which have the base sequence H$_2$N-(CH$_2$)$_6$-5'-AGG ACT TCA CCG TAT TGC-3'. The DNAs are custom designed and obtained from Sigma Aldrich. 2 $\mu$M of probe DNAs, prepared in PBS, is flowed through the microfluidic channel over the SLG channels. The device is left for immobilisation with probe DNAs overnight at 4 $^\circ$C inside a wet chamber following the recipe given in \cite{campos2019attomolar}. The amine group of the pDNA reacts with the succinimide group of PBASE through conjugation reaction (Fig. \ref{fig:functionalization}(c)). The excess pDNA is then removed from the channel with PBS rinsing. Note that although the ssDNAs and the double-stranded DNAs (dsDNAs) shown in Figs. \ref{fig:functionalization}(c)-(d) are depicted as vertically aligned over the linker molecules, the orientation of DNAs tethered to surfaces through their single end can be influenced by the electrical potential of the surface, electrolyte flow conditions, the length of the DNAs, temperature, pH, ionic strength of the electrolyte, and the existence of the linker molecules. It is shown through molecular dynamics simulations that under zero potential of the surface and in the absence of the solution gate potential and the linker molecules, the flexible nature of the ssDNAs results in tilted and near-parallel orientation on the surface \cite{kabelavc2012influence}. On the other hand, dsDNAs attain more vertical alignment under zero potential due to their higher rigidity \cite{kabelavc2012influence}. However, to the best of author's knowledge, the effect of the solution gate potential and the linker molecules has not been studied in the literature.

After pDNA immobilisation, the passivation of the unbound PBASE molecules is necessary to prevent nonspecific binding of target DNAs (tDNAs). This is performed by flowing 100 mM ethanolamine (NH$_2$CH$_2$CH$_2$OH) solution prepared in DI water through the microfluidic channel. Ethanolamine reacts with amine-reactive succinimide group of unbound PBASE molecules. With the passivation of PBASE, the device becomes ready for sensing and communication experiments with tDNAs.

\section{Electrical Characterisation of MC Receiver}
For the electrical characterisation of the fabricated devices, direct-current (DC) measurements are taken using a high-precision source measure unit (SMU, Keysight B2902A), which is connected to the device electrodes via high-impedance passive probes, as shown in Fig. \ref{fig:microfluidic_setup2}. On the other hand, the mobility of the GFET channels is measured before functionalisation in a back-gate configuration using EverBeing probe station. Based on the linear approximation of the transfer curve, the mobility is calculated as $(240.62 \pm 23.47)$ cm$^2$/V$\cdot$s. 

\subsection{Transfer Characteristics}
After each step of functionalisation, transfer characteristics of the devices are obtained with a constant drain-to-source bias $V_{ds} = 100$ mV, and a solution gate potential $V_g$ varying between -0.2 V and 1.2 V. The sweep rate of $V_g$ is set to 140 mV/s. All data are obtained after removal of excessive functional molecules from the microfluidic channel and the SLG surfaces by rinsing with PBS. This ensures that no change occurs in transfer characteristics due to ongoing chemical reactions. The PBS (pH 7.4) is used as the electrolyte in all measurements of transfer characteristics. 
\begin{figure*}[!t]
	\centering
	\subfigure[]{
		\includegraphics[width=7.5cm]{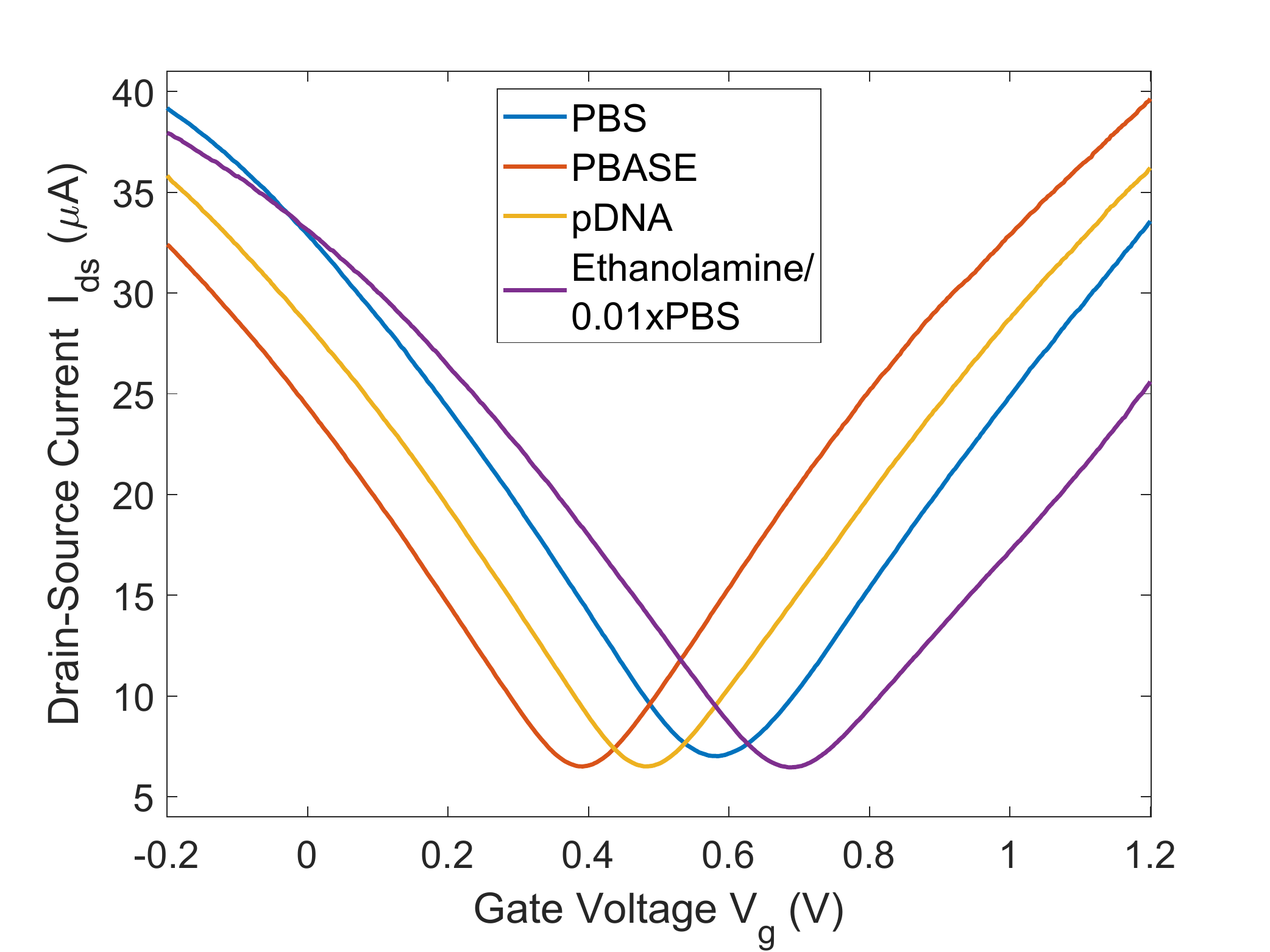}}
	\subfigure[]{
		\includegraphics[width=7.5cm]{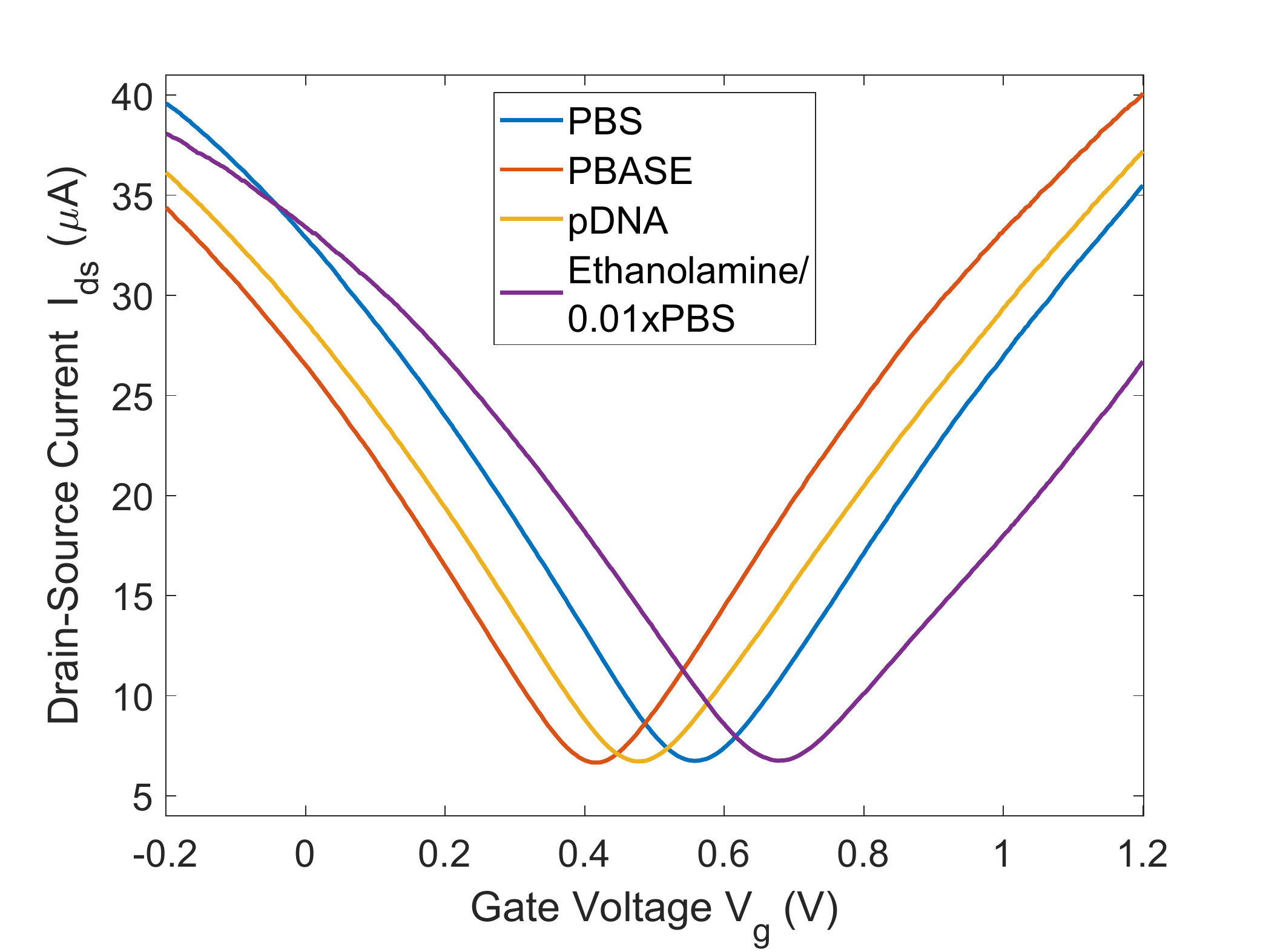}}
	\subfigure[]{
		\includegraphics[width=7.5cm]{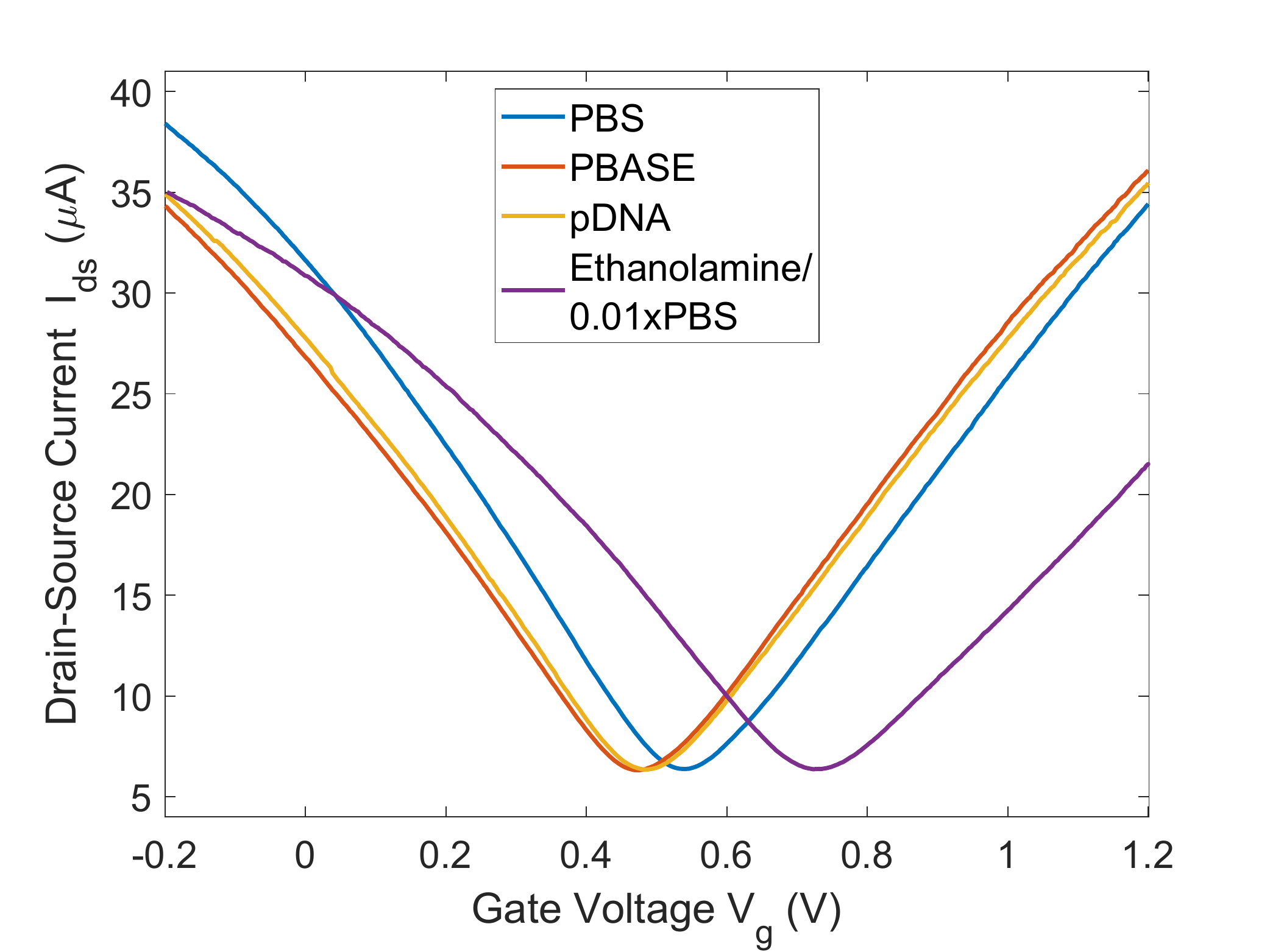}}		
	\subfigure[]{
		\includegraphics[width=7.5cm]{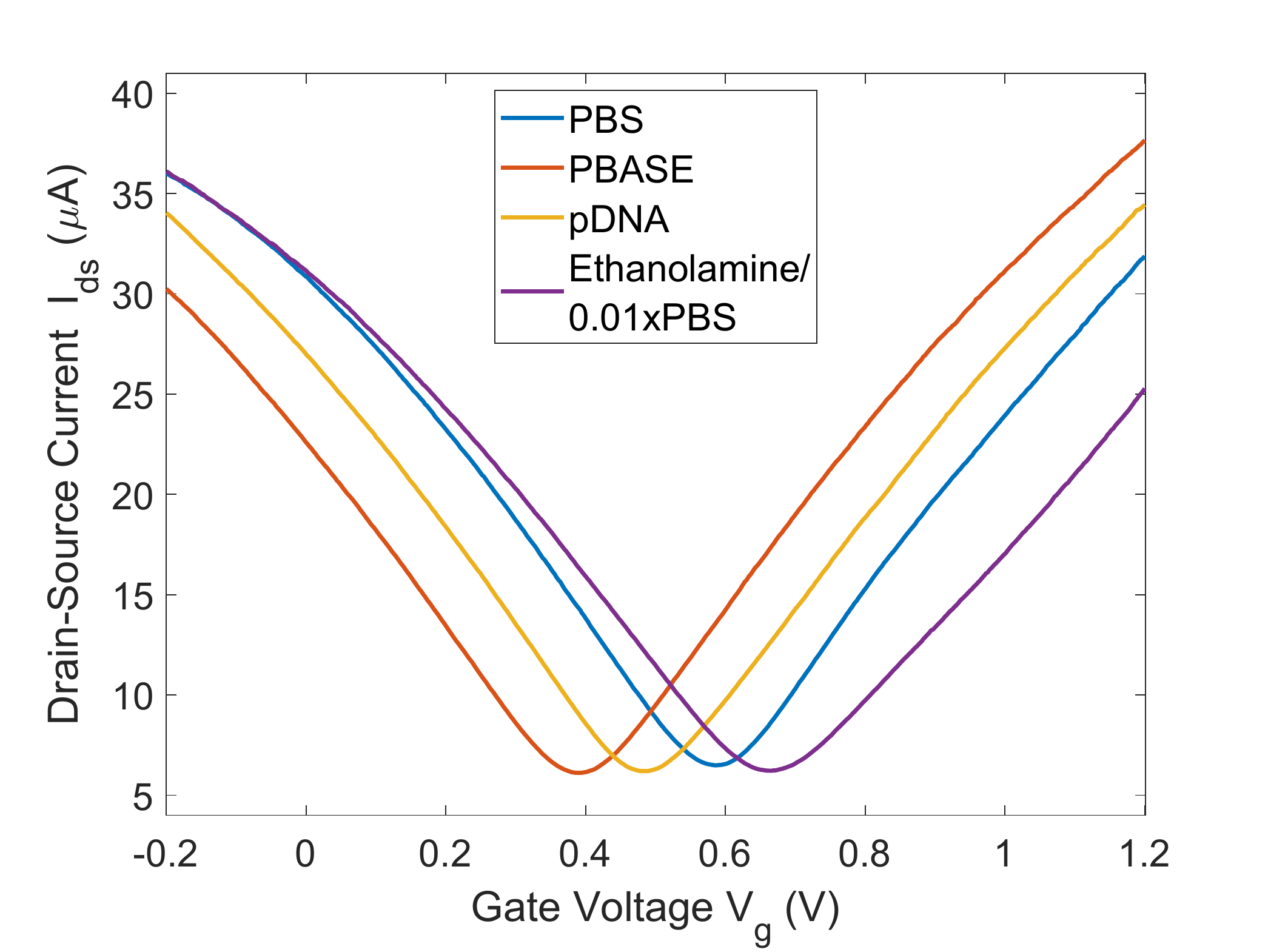}}
	\caption{Transfer characteristics of four channels at different steps of functionalisation in terms of drain-source current $I_{ds}$ as a function of varying gate voltage $V_g$ with sweep rate 140 mV/s. In all measurements drain-source voltage is held constant at $V_{ds} = 0.1$ V. }
	\label{fig:transfer}
\end{figure*}

The measurements are taken from four of the SLG channels in the MC receiver, and results are provided in Fig. \ref{fig:transfer}. Hysteresis was negligible for all channels (see Fig. \ref{fig:hysteresis} in Appendix \ref{AppendixB}), thus, only the forward sweep of $V_g$ is demonstrated. First measurement is taken with only the PBS electrolyte inside the microfluidic channel prior to the functionalisation process. The p-type behaviour and ambipolar characteristics of the SLG-based devices are revealed with the charge neutrality point (CNP), i.e., the gate voltage of the minimum conductance, observed at $\sim$0.57 V on average over four channels with a standard deviation of $\sim$0.02 V. Upon functionalisation with PBASE linker, a negative shift of the CNP by $150 \pm 61$ mV is observed, indicating n-type doping. The negative shift of the CNP after the PBASE functionalisation was previously attributed to the dominance of the n-type doping effect of DMF in competition with the p-type doping effect of the PBASE molecules over long incubation times \cite{wu2017doping,ghosh2018selective}. Note that the large standard deviation of the CNP shift is mostly resulting from the atypical behaviour observed in the third GFET channel, the transfer characteristics of which are demonstrated in Fig. \ref{fig:transfer}(c). In this channel, only 67 mV-shift in CNP  is observed with the functionalisation of PBASE. The significantly smaller shift compared to other channels, which manifest consistent transfer characteristics, can be indicative of the poor functionalisation of the PBASE linkers on this particular GFET channel. The poor functionalisation can be due to the residual polymers or insulator material on the graphene surface remaining from the fabrication process and preventing the non-covalent attachment of the PBASE molecules.

On the other hand, the immobilisation of pDNAs resulted in a positive shift of the CNP in consistence with the previous literature \cite{xu2017real}. DNA molecules are negatively charged at pH 7.4, attracting hole carriers to the graphene surface, thus, contributing to the p-type doping \cite{xu2017real}. The shift of the CNP is observed as $66 \pm 41$ mV over the four channels. Again, the large standard deviation can be attributed to the third channel given in Fig. \ref{fig:transfer}(c), where the CNP shift is only 11 mV. This is again indicative of the poor functionalisation of the PBASE molecules, which in turn results in a very low concentration of immobilised pDNAs. Other channels, on the other hand, show similar CNP shifts, indicating more consistent immobilisation of DNAs. 

The last measurements are taken after the passivation of excess PBASE linkers with the ethanolamine, and the introduction of the 0.01xPBS to the microfluidic channel to be used for the following sensing and communication experiments. 0.01xPBS is the 100-fold diluted version of PBS with DI water (18.2 M$\Omega$$\cdot$cm). While the ethanolamine does not possess any charge, the observed positive shift of the CNP is consistent with the previous literature reporting increased p-type doping with decreasing ionic concentration of the buffer solution \cite{chen2013label}. Also note that in all of the measured GFET channels, a mobility reduction is observed upon passivation with ethanolamine in 0.01xPBS. The reason for switching to the diluted version of PBS for sensing and communication experiments is to decrease the effect of the Debye screening for enhancing the sensitivity of the device for the hybridisation events on the SLG surface. Note that in all measurements, leakage current $I_{gs}$ has been detected to be under 15 nA (see Fig. \ref{fig:leakage} in Appendix \ref{AppendixB}), and therefore, its effect on the transfer curve characteristics is negligible. 

With the help of the transfer characteristics, we can deepen our analysis by determining the surface density of the immobilised pDNAs. To this end, we need to first determine the electrolyte gate capacitance, which can be approximated by the overall capacitance of three parallel plate capacitors connected in series:  
\begin{align}
	C_G = \left(\frac{1}{C_{Gr}} + \frac{1}{C_{Pt}} + \frac{1}{C_Q} \right)^{-1}, \label{capacitance}
\end{align}
where $C_{Gr}$ is the EDLC between the graphene and the electrolyte, $C_{Pt}$ is the EDLC between Pt gate electrode and electrolyte, and $C_Q$ is the quantum capacitance of graphene \cite{xu2017real}. The EDLC of graphene to electrolyte can be calculated as $C_{Gr} = A_{Gr}  \epsilon_r \epsilon_0/\lambda_D$, with $A_{Gr} = 40$ $\mu$m$\times100$ $\mu$m $= 4\times10^3$ $\mu$m$^2$ being the area of graphene surface exposed to electrolyte,  $\epsilon_0$ is the vacuum permittivity, and $\epsilon_r$ is the relative permittivity of PBS electrolyte, which is only slightly lower than the one of water, thus, taken as $\epsilon_r \approx 80$ \cite{xu2017real,zheng2013electrical}. Lastly, $\lambda_D$ is the Debye length which gives the thickness of the EDLC, and it can be approximated in aqueous solutions as $\lambda_D \approx 0.3/\sqrt{\rho_{ion}}$ in nm, with $\rho_{ion}$ being the ionic density in M \cite{goldsmith2019digital}. For 1xPBS buffer, the ionic density is $\sim$150 mM, thus, $\lambda_D \approx 0.77$ nm. The resulting EDLC for graphene is $C_{Gr} \approx 3.68$ nF. The EDLC between the Pt electrode and the electrolyte can be obtained similarly as $C_{Pt} = A_{Pt}  \epsilon_r \epsilon_0/\lambda_D$. However, since the surface area of the Pt wire inside the electrolyte ($A_{Pt} = l_{Pt} d_{Pt} \pi/2  = 1$ cm$\times 0.5$ mm$\times \pi/2 = 7.85 \times 10^6$ $\mu$m$^2$, calculated as the area of an half sphere of 1cm-length and 0.5mm-diameter) is significantly larger than the graphene surface area ($A_{gr} = 4\times10^3$ $\mu$m$^2$), $C_{Pt} $ can be neglected. Lastly, the quantum capacitance of graphene per unit area has been reported as $c_q \approx 2$ $\mu$F$\cdot$cm$^{-2}$ \cite{xu2017real, meric2008current}, which gives $C_Q = c_q \times A_{gr} = 8 \times 10^{-2}$ nF. The overall gate capacitance given by \eqref{capacitance} then becomes $C_G \approx 7.83\times10^{-2}$ nF. 

The effective electric charge of a single immobilised 18-mer pDNA screened by the EDL can be written as
\begin{align}
	q_{pDNA}  = 18 \times q_e \times e^{-r/\lambda_D} , \label{effectivecharge}
\end{align}
where $q_e$ is the elementary charge, $r$ is the effective length of the pDNA taken as the half of its length, i.e., $r = 18$(basepairs)$\times0.34$(nm/basepair)$\times 1/2 = 3.06$ nm, by assuming a vertical orientation for the single-stranded pDNAs following the analysis in \cite{xu2017real}, where similar solution gate potentials and the same type of linker molecules are used. 

Finally, the surface density of the immobilised pDNAs can be written as a function of the average shift in $V_{CNP}$ upon pDNA functionalisation: 
\begin{align}
	n_{pDNA}= \frac{\Delta V_{CNP} C_G}{q_{pDNA} A_{gr}}, \label{pDNAdensity}
\end{align}
which gives $n_{pDNA} \approx 2\times10^3$ $\mu$m$^{-2}$. A similar surface density ($\sim1.14\times10^3$ $\mu$m$^{-2}$) for pDNA was previously reported in \cite{xu2017real}.

\subsection{Sensing Response}
\begin{figure*}[!t]
	\centering
	\subfigure[]{
		\includegraphics[width=8cm]{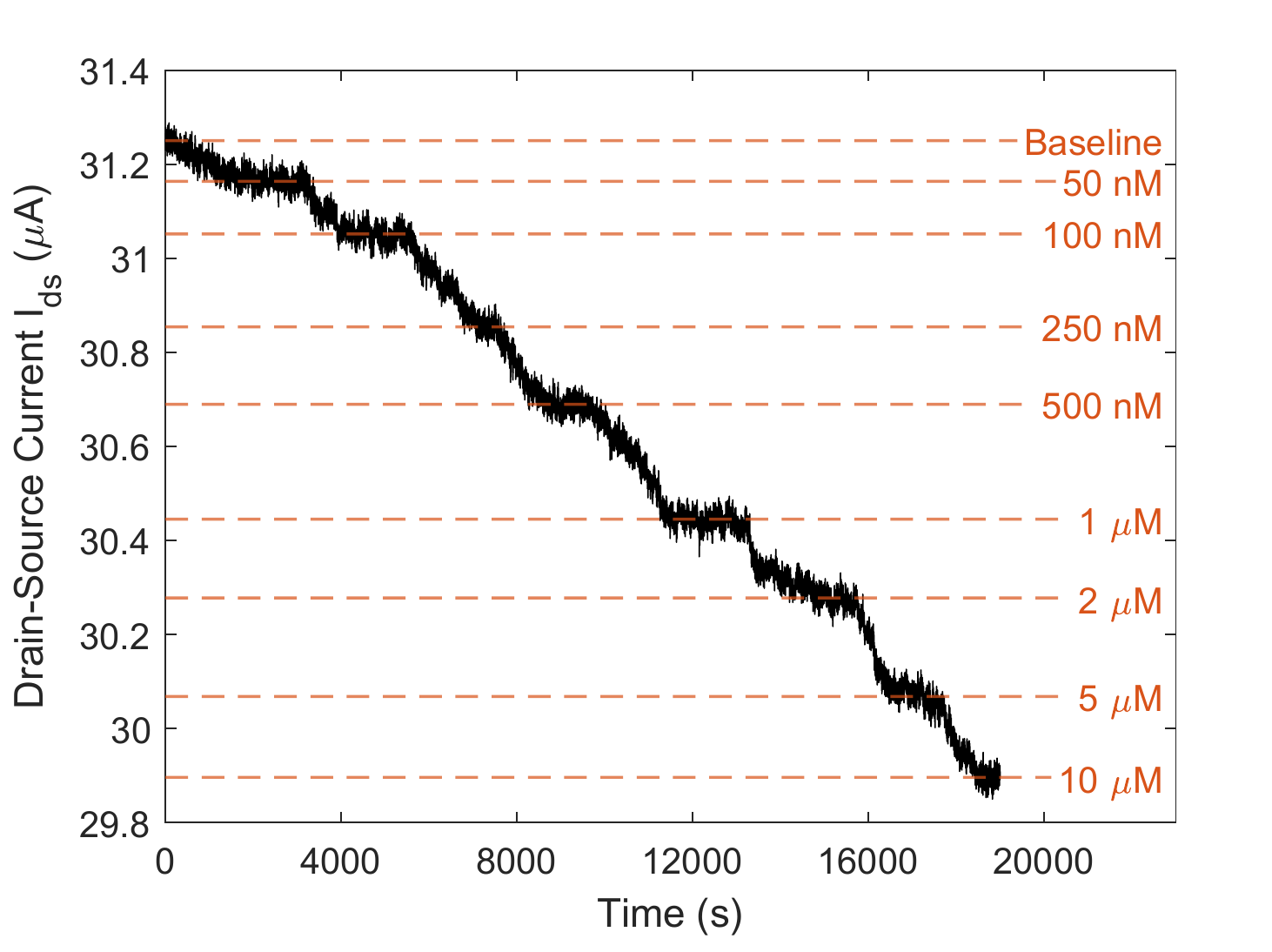}}	
	\subfigure[]{
		\includegraphics[width=8cm]{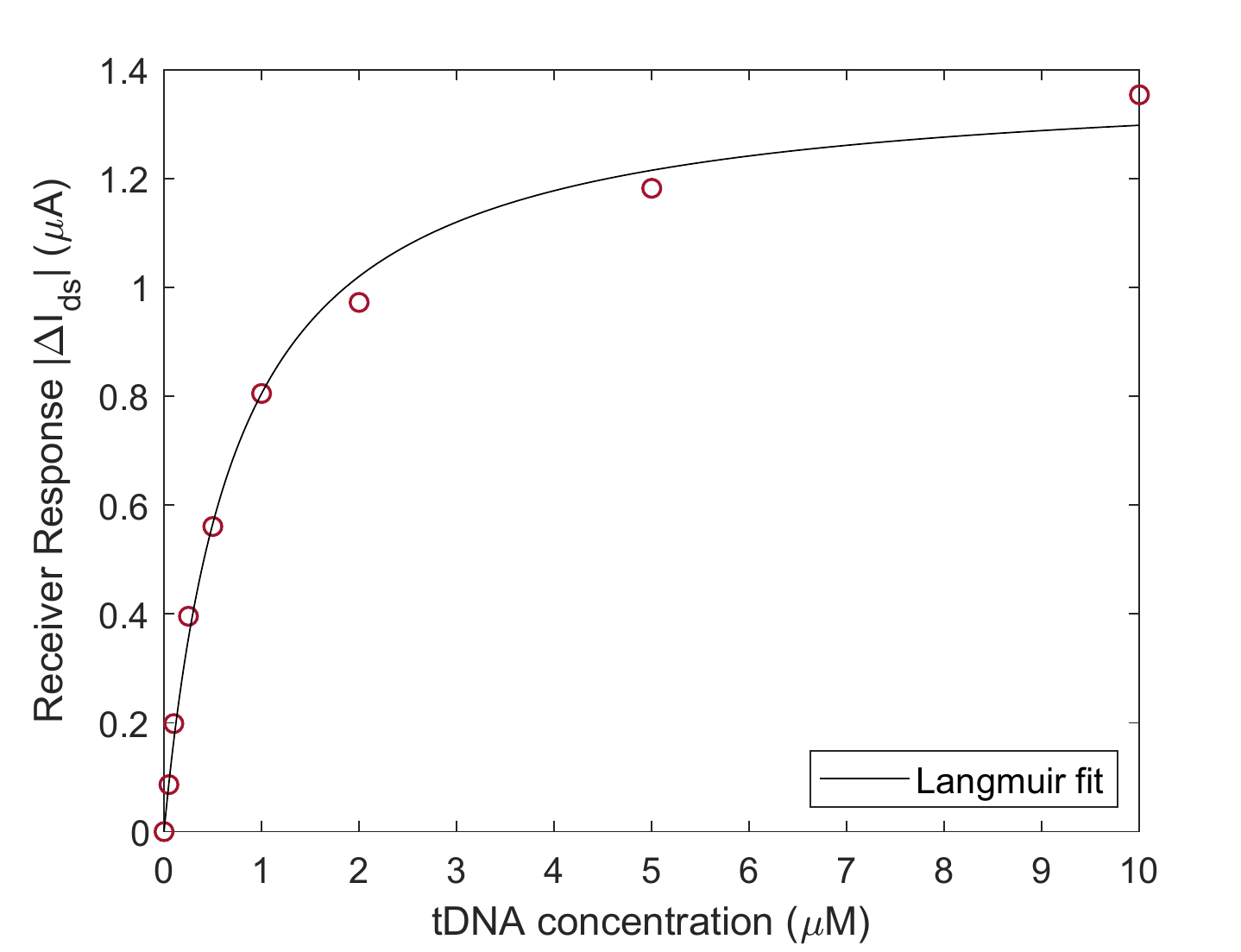}}
	\caption{(a) Real-time sensing response of the MC receiver in terms of drain-source current $I_{ds}$ with varying concentration of complementary target DNAs (tDNAs). (b) Equilibrium sensing response fitted by the Langmuir adsorption isotherm. Resulting dissociation constant for pDNA-tDNA hybridisation is $K_D = 730$ nM.}
	\label{fig:sensogram}
\end{figure*}

For determining the sensing characteristics of the MC receiver, complementary 18-mer target DNA (tDNA: 5'-GCA ATA CGG TGA AGT CCT-3', obtained from Sigma Aldrich) is prepared in 0.01xPBS solution. tDNAs of varying concentrations (50 nM, 100 nM, $\dots$ 10 $\mu$M) are successively flowed through the microfluidic channel in the order of increasing concentration, and $I_{ds}$ is recorded in real time with $V_{ds} = 100$ mV and $V_g = 0$ V. During the experiment, the volumetric flow rate is held constant at $u_V = 80$ $\mu$l/min. The results of the measurements are provided in  Fig. \ref{fig:sensogram}(a), where a decrease in the drain-source current is observed with increasing tDNA concentration, implying n-type doping effect in contrast to the p-type doping of pDNAs. The n-type doping effect upon target DNA hybridisation or probe DNA immobilisation was previously reported in \cite{campos2019attomolar,manoharan2017simplified,dong2010electrical,yue2017electricity,xu2014electrophoretic,ghosh2018selective}, where the effect is mainly attributed to the partial interaction of the DNAs with the graphene surface through $\pi-\pi$ stacking of the nucleobase aromatic rings resulting in direct electron transfer to graphene instead of electrostatic gating. The electron transfer from DNA upon immobilisation was also reported for CNT transistors \cite{star2006label}. Moreover, it is known that DNA molecules immobilised on a surface with their single end can be stretched in parallel to the surface under lateral flow \cite{van2008see}, and the extent of this conformational change can be increased by a positive surface potential attracting the negatively charged DNA molecules to the surface \cite{xu2014electrophoretic,kabelavc2012influence}. Note that in our case, the graphene surface is continuously exposed to a positive drain-source bias during the sensing measurements. Therefore, we speculate that the conformational change resulting from microfluidic flow, hybridisation, and the positive surface charge of graphene brings the hybridised DNA molecules closer to the graphene surface, causing their partial interaction and electron transfer. Also, compared to the measurements taken with the single-stranded pDNAs under no-flow conditions, the ionic strength of the electrolyte (0.01$\times$PBS), in which the sensing experiments are performed, is significantly lower (compared to 1$\times$PBS), such that the attractive electrostatic force caused by the positive surface potential extends more into the electrolyte without being significantly screened \cite{kaiser2010conformations}. Given that the hybridised dsDNAs carry twice the amount of negative charge of the ssDNAs, it can be considered that the hybridised pDNA-tDNA pairs, in our case, are more strongly attracted to the graphene surface compared to pDNAs \cite{rant2004dynamic}. The stronger electrostatic attraction combined with the stretching effect of lateral microfluidic flow supports our argument. However, this requires further confirmation, potentially through molecular dynamics simulations of both ssDNAs and hybridised dsDNAs under similar conditions to understand the effect of lateral flow, surface potential, solution-gate potential, and the linker molecules. 

Each working concentration of tDNAs were propagated in the channel until $I_{ds}$ reaches a plateau. The value of $I_{ds}$ at these plateaus are used to construct the sensing response graph of the MC receiver, which is provided in Fig. \ref{fig:sensogram}(b). The response curve is fitted by the Langmuir adsorption isotherm, i.e.,
\begin{align}
	\Delta I_{ds}/\Delta I_{ds,sat} = \frac{1}{1 + K_D / C_{tDNA} }, \label{langmuir}
\end{align}
where $\Delta I_{ds,sat}$ is the receiver response in saturation, which occurs when all the probe DNAs are hybridised. $K_D$ is the dissociation constant of pDNA-tDNA hybridisation, and $C_{tDNA}$ is the applied concentration of tDNAs. The curve fitting gives the dissociation constant as $K_D = 730$ nM for the DNA hybridisation on the fabricated MC receiver, and the receiver response at saturation as $\Delta I_{ds,sat} = 1.393$ $\mu$A. 
\begin{figure}[!t]
	\centering
	\subfigure[]{
		\includegraphics[width=8cm]{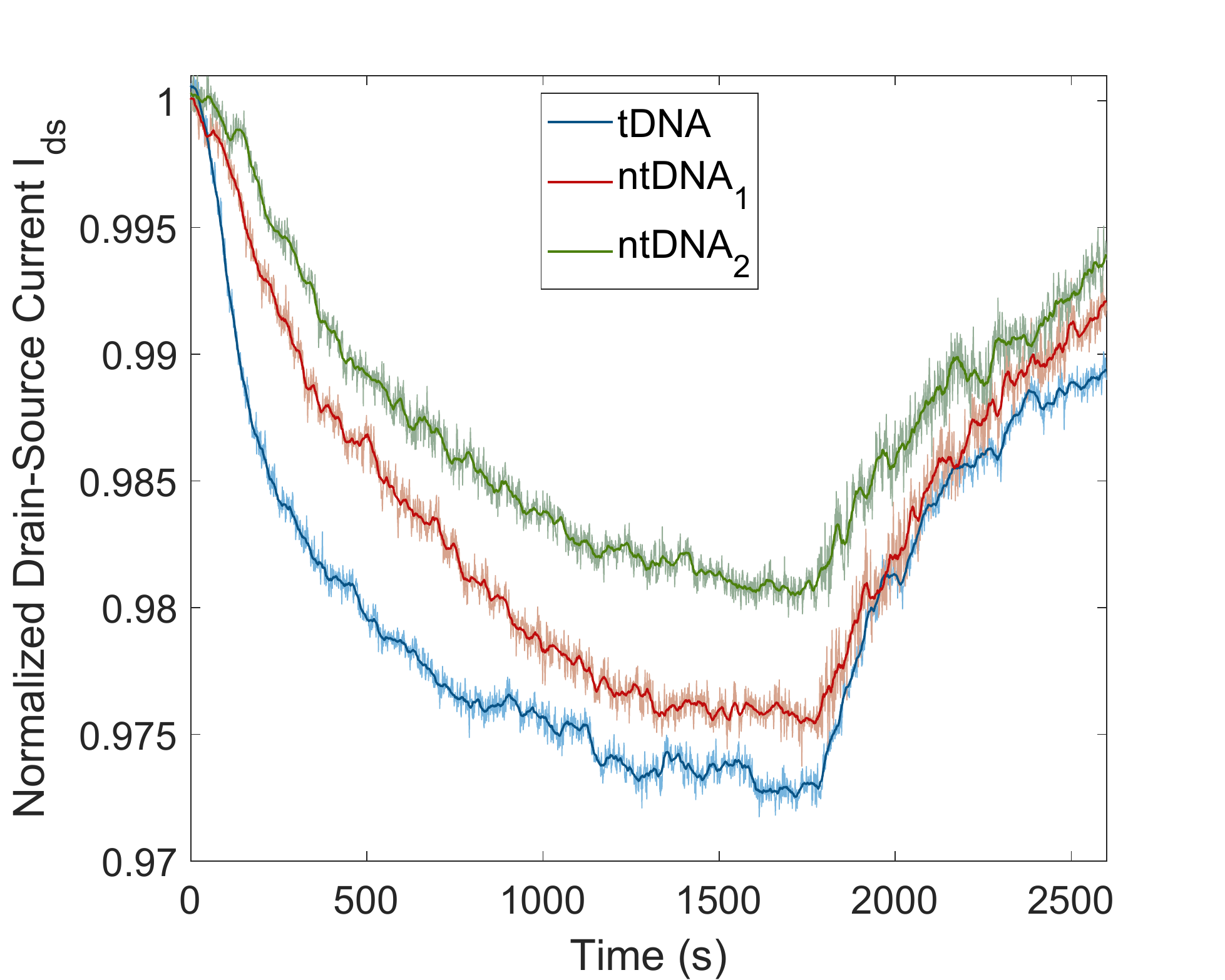}}	
	\subfigure[]{
		\includegraphics[width=8cm]{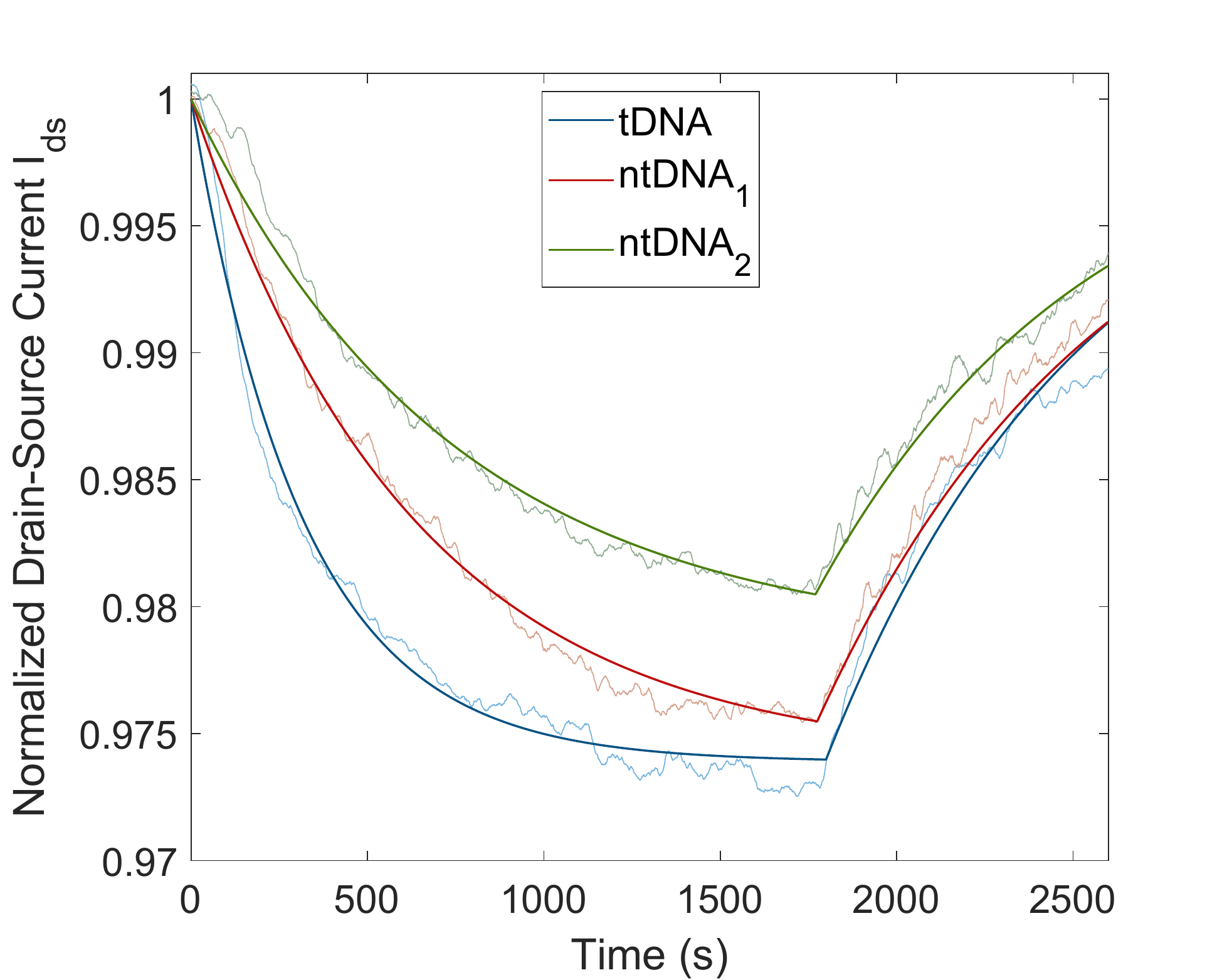}}
	\caption{Specificity analysis of the MC receiver. (a) Real-time sensing response for complementary tDNA, non-complementary ntDNA$_1$ with single base-pair mismatch, and non-complementary ntDNA$_2$ with 7 base-pair mismatches (see Table \ref{table:kinetic}). At $t \approx 1800$ s, the DNA solutions are replaced with 0.01xPBS solution to allow dissociation of the hybridised DNAs. (b) Real-time sensing response fitted by the Langmuir adsorption/desorption model, equations \eqref{langmuir_assoc}-\eqref{langmuir_dissoc}.}
	\label{fig:interference}
\end{figure}

\subsection{Specificity}
The specificity of the MC receiver against the complementary tDNAs is evaluated by comparing the receiver's response to different ssDNAs, which are not complementary to the pDNAs. The ultimate specificity can be determined with the application of an ssDNA having only one single base-pair mismatch. For this, we use 18-mer ntDNA$_1$ with the base sequence 5'-GCA ATA CGG CGA AGT CCT-3', which has the mismatch in its 10$^{th}$ base pair, where T to C mutation occurs. Another test is performed with the application of 18-mer ntDNA$_2$ (5'-GCA CGT CGG CGT CGT CAT-3'), which has 7 base-pair mismatches. Complementary tDNA is also applied for comparison. All DNAs are dissolved in 0.01xPBS  with 1 $\mu$M working concentration. The measurement results before and after a moving mean filter of 21-second window length is applied are provided in Fig. \ref{fig:interference}(a).

\begin{table*}[!t]\small
	\centering
	\begin{threeparttable}
		\centering
		\caption[Kinetic constants of DNA hybridisation measured by MC receiver]{Kinetic constants of DNA hybridisation measured by MC receiver}
		\label{table:kinetic}
		\begin{tabular}{ccccc}
			\toprule	
			\textbf{DNA}   & \textbf{Sequence}   & \textbf{Binding} & \textbf{Unbinding} & \textbf{Dissociation}\\ 
			\textbf{}   & & \textbf{rate $k^+$} & \textbf{rate $k^-$} &\textbf{constant $K_D$} \\ 
			\textbf{}   & & \textbf{(M$^{-1}$s$^{-1}$)}  & \textbf{($\times~10^{-4}$ s$^{-1}$)}  &\textbf{($\times~10^{-6}$ M)} \\ 
			\toprule
			\textbf{tDNA}  & 5'-GCA ATA CGG TGA AGT CCT-3' & 1814.9 & 13.538&  0.746  \\ \midrule
			\textbf{ntDNA$_1$}  &  5'-GCA ATA CGG \textcolor{red}{C}GA AGT CCT-3'& 355.3   & 12.454 &  3.506 \\ \midrule
			\textbf{ntDNA$_2$}    & 5'-GCA \textcolor{red}{CGT} CGG \textcolor{red}{C}G\textcolor{red}{T} \textcolor{red}{C}GT C\textcolor{red}{A}T-3' & 48.9 & 13.110&  26.829 \\ \bottomrule
		\end{tabular}%
	\end{threeparttable}	
\end{table*}%

The response curve of the MC receiver for different DNAs is fitted by the Langmuir model of adsorption to determine the kinetic rates of the DNA hybridisation for the three DNA sequences. The solution of the Langmuir model gives the time-varying response of the MC receiver during the association and dissociation phases of DNA hybridisation \cite{xu2017real} as follows 
\begin{align}
	\Delta I_{ds} (t) = \Delta I_{ds,eq} \bigl(1-e^{-(k^+ c_{in} + k^-)t}\bigr), \text{for}~~0 \le t \le t_d, \label{langmuir_assoc}
\end{align}
\begin{align}
	\Delta I_{ds} (t) = \Delta I_{ds}(t_{d}) e^{-k^- t}, \text{for}~~t > t_d, \label{langmuir_dissoc}
\end{align}
where, $c_{in}$ is the input concentration, which is set to $c_{in} = 1$ $\mu$M for all DNAs. $t_d$ denotes the time of dissociation, and $\Delta I_{ds,eq}$ is the asymptotic value of the sensing response, which occurs when the hybridisation reaches equilibrium. The variables to be fitted are the binding rate $k^+$, unbinding rate $k^-$, and $\Delta I_{ds,eq}$. The fitted response is plotted in Fig. \ref{fig:interference}(b), and the resulting kinetic rates are provided in Table \ref{table:kinetic}, which shows that the binding rate of the target DNAs substantially decreases with increasing number of base-pair mismatches.

The nonlinear curve fitting gives the dissociation constant of the tDNA as $K_D(\text{tDNA}) = k^-(\text{tDNA})/k^+(\text{tDNA}) \approx 746$ nM, which is very close to the value obtained by the fitting of the sensor response, i.e., 730 nM. On the other hand, we obtain higher dissociation constants for non-complementary DNAs, i.e., $K_D(\text{ntDNA}_1) = 3.506$ $\mu$M and $K_D(\text{ntDNA}_2) = 26.829$ $\mu$M, indicating the specificity of the MC receiver against the complementary tDNAs. 

\section{Communication Performance}

Time-varying communication experiments are performed in the microfluidic testbed to reveal the detection performance of the MC receiver. For these experiments, both inlets are utilised as shown in Fig. \ref{fig:microfluidic_setup2}. One of the inlets is connected to the reservoir containing the 0.01xPBS buffer solution, and the other one is connected to the reservoir containing the tDNA solution in 0.01xPBS buffer. Manually controlled mechanical switches are utilised as they proved more effective in stopping the fluid flow into the channel than the digital control due to the fact that the pressure controller can drift out of calibration as the experiments progress. Accordingly, for the transmission of tDNAs, the buffer flow is rapidly stopped through the mechanical switch along the buffer line, and the tDNA line is opened at the same time. When the transmission ends, flow in the tDNA line is stopped, and the buffer flow is simultaneously started again by means of mechanical switching.

\subsection{Time-varying Response}
\begin{figure}[!t]
	\centering
	\subfigure[]{
		\includegraphics[width=8cm]{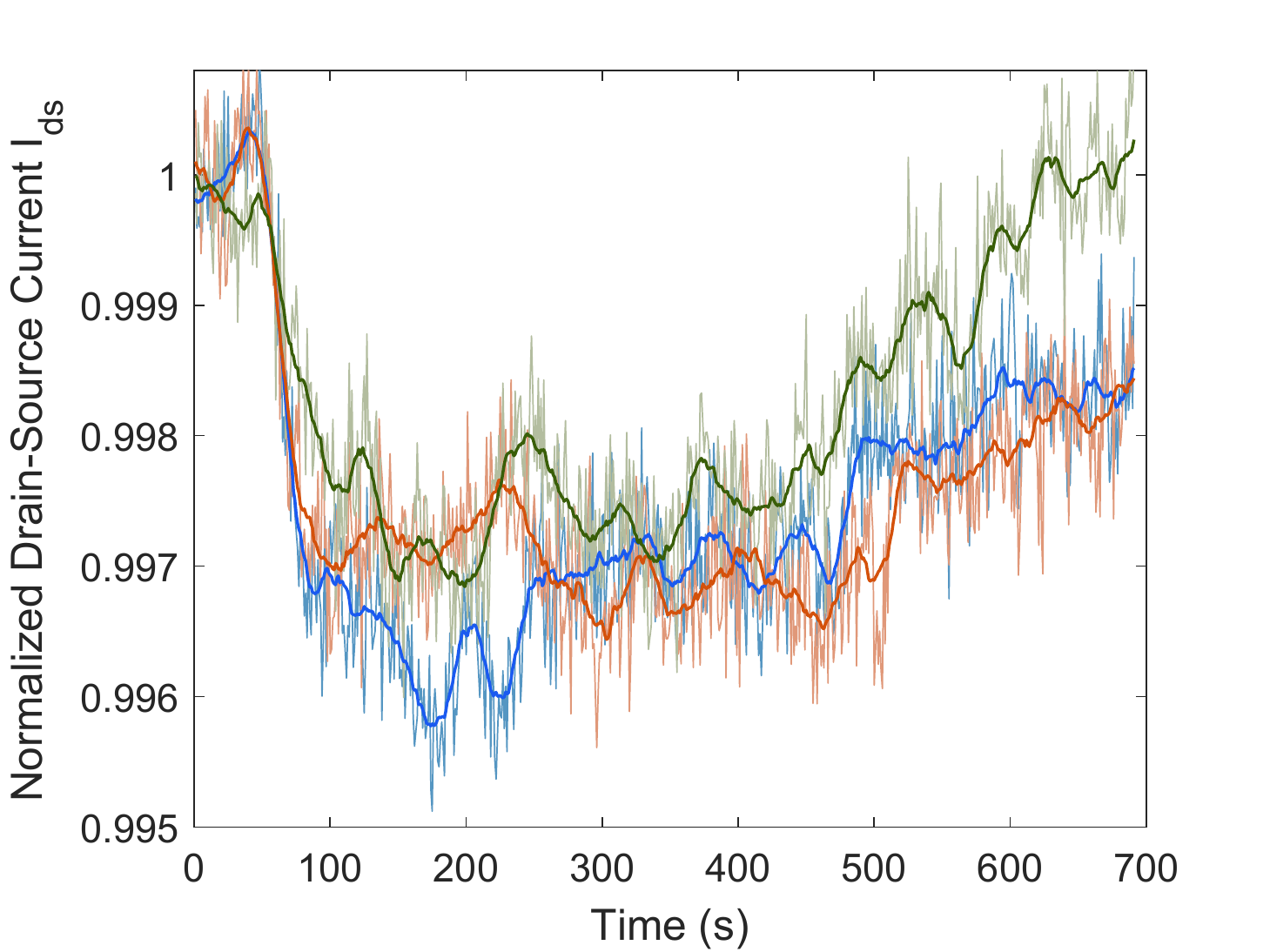}}	
	\subfigure[]{
		\includegraphics[width=8cm]{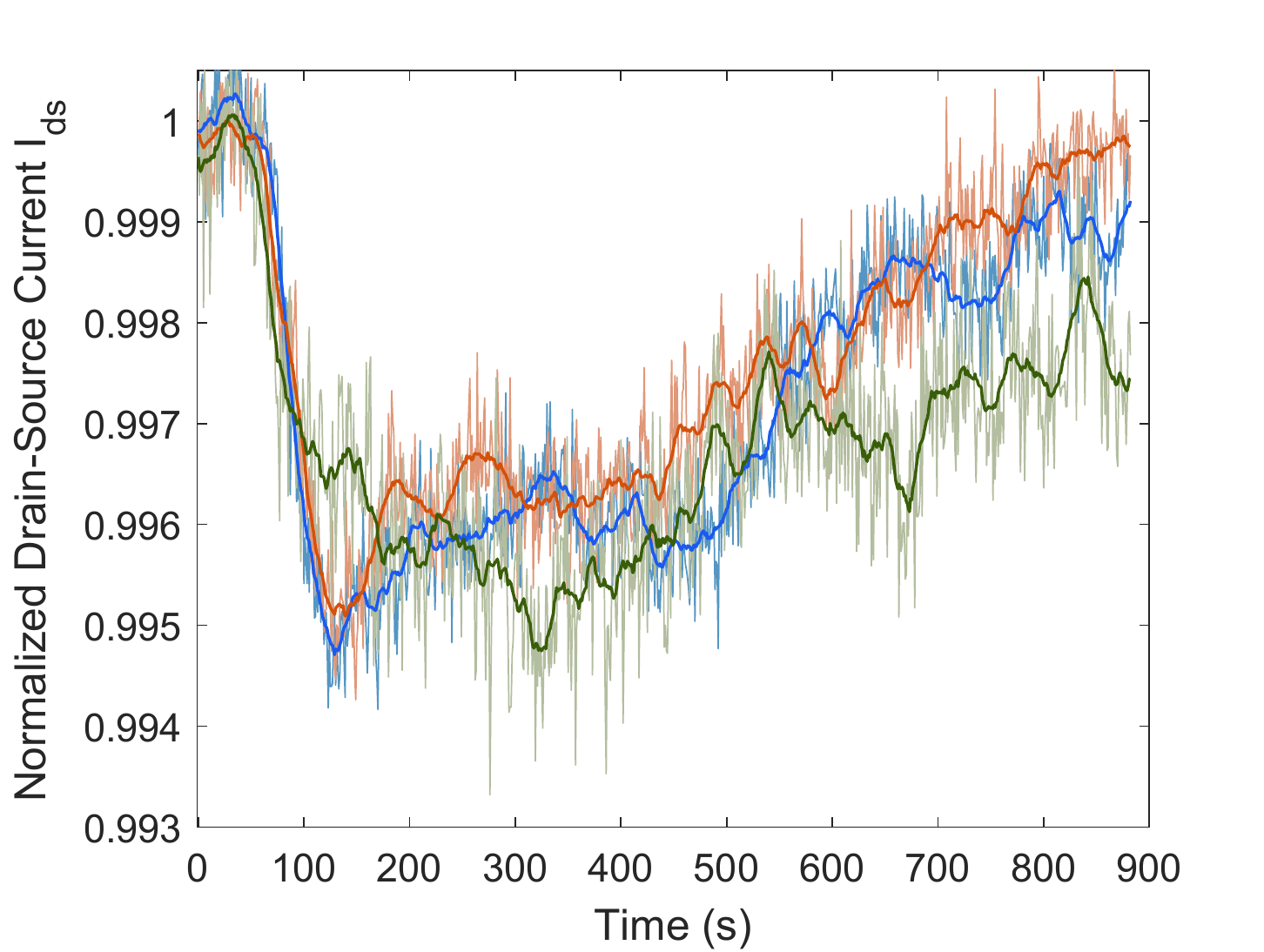}}
	\subfigure[]{
		\includegraphics[width=8cm]{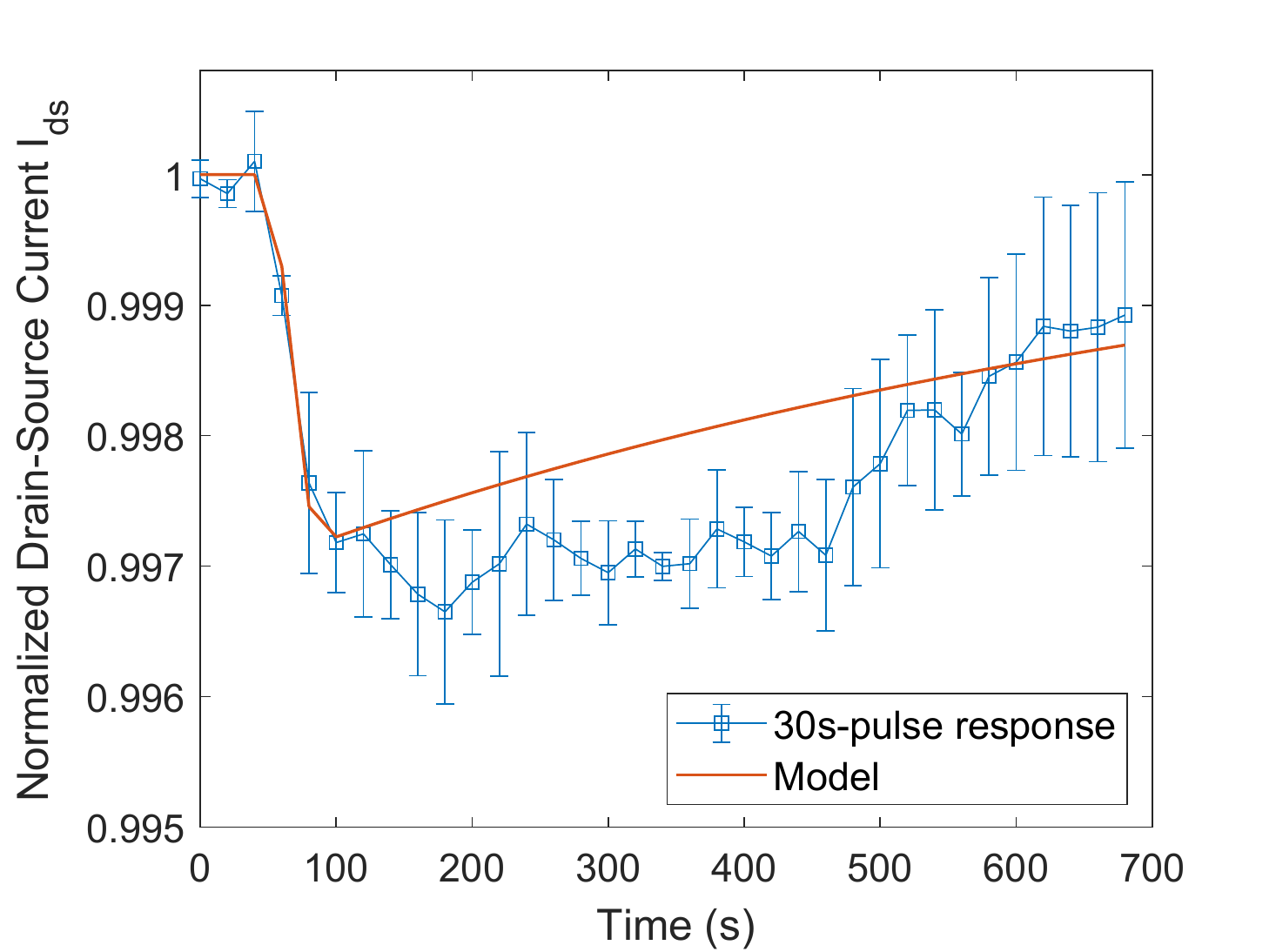}}	
	\subfigure[]{
		\includegraphics[width=8cm]{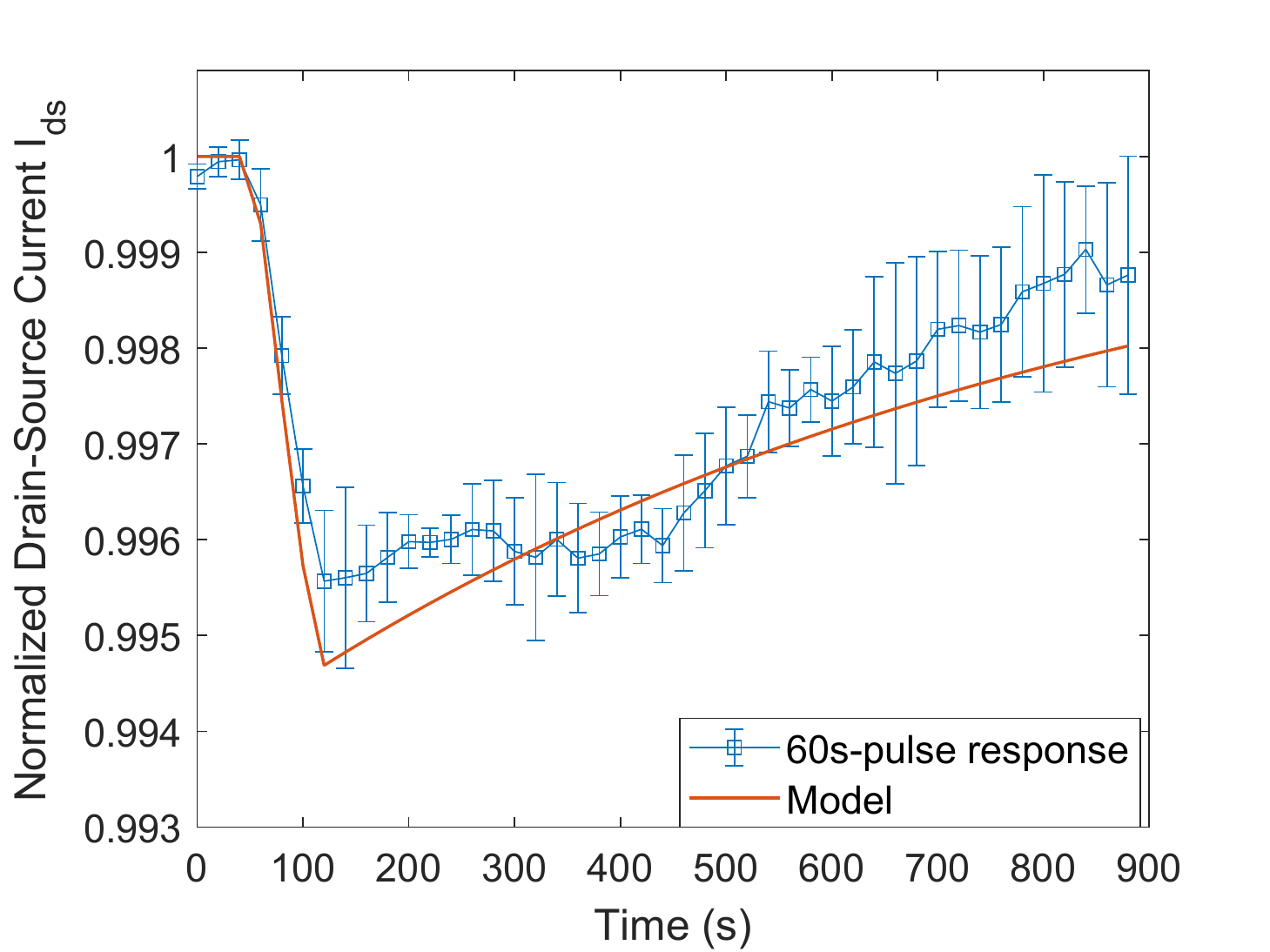}}	
	\caption{Normalised pulse response of the MC receiver in terms of drain-source current $I_{ds}$ with constant operating voltages set to $V_g = 0$ V and $V_{ds} = 0.1$ V. (a, b) Three independent measurements taken from the same channel for 30-second-long and 60-second-long 1 $\mu$M tDNA pulses, respectively. (c, d) Pulse responses fitted by the normalised output of the transformed model given in \eqref{eq:R3}. }
	\label{fig:pulsefitnew}
\end{figure}

To determine the time-varying response of the receiver, varying length pulses of $1$ $\mu$M tDNAs are flowed through the microfluidic channel. The results of three independent measurements taken from the same channel are provided in Fig. \ref{fig:pulsefitnew}(a) and Fig. \ref{fig:pulsefitnew}(b) for 30-second and 60-second pulses, respectively. As with the previous cases, a 21-second-length moving mean filter is applied for each measurement.

The time-varying response of the MC receiver to finite-length concentration pulses can be described by the analytical microfluidic MC model developed in \cite{kuscu2018modeling}. This approximate model gives the receiver response in terms of number of bound receptors, i.e., 
\begin{align} 
	\label{eq:R2}
	N_R(t) =& N_{R,eq} \left( 1- \frac{\mathcal{W}_0\bigl[\alpha^\ast \exp \left(\alpha^\ast - \beta^\ast (t-t_a)\right)\bigr]}{\alpha^\ast}\right) \Big( \Theta\left[t-t_a\right] - \Theta\left[t-t_d-\epsilon \right] \Big) \\ \nonumber
	&- \gamma^\ast \mathcal{W}_0 \Bigg[ - \frac{N_{R,0}}{\gamma^\ast} \exp \left( \frac{-k^+ N_{R,0} - k_T^\ast k^- (t-t_d)}{k^+ \gamma^\ast}\right)\Bigg] \Theta\left[t-t_d-\epsilon \right],
\end{align}
where $t_a$ and $t_d$ denote the start times of association (i.e., pDNA-tDNA hybridisation in this case) and dissociation phases, respectively, $N_R(t)$ is the number of bound receptors at time $t$, $N_{R,eq}$ is the number of bound receptors at equilibrium, $N_{R,0} = N_R(t_d)$ is the number of bound receptors when the dissociation starts. Here $\mathcal{W}_0[.]$ denotes the principal branch of the Lambert $\mathcal{W}$ function, and $\Theta[.]$ denotes Heaviside step function with $\Theta[0] = 1$. The parameters $\alpha^\ast$, $\beta^\ast$ and $\gamma^\ast$ are provided in [51, Eqs. (37-39)] as functions of $N_{R,max}$, which is the upper limit of the receiver response in terms of bound number of receptors. In theory, $N_{R,max}$ can be taken as equal to the total number of receptors. Lastly, the transport parameter $k_T^\ast$ incorporating the effect of the microfluidic channel geometry and flow velocity on the molecular transport dynamics is also provided in [51, Eq. (13)] . 

To use this model for fitting the electrical measurements obtained by the graphene-based MC receiver, we need to transform the variables in \eqref{eq:R2} to obtain a function of $\Delta I_{ds}$. The number of bound receptors is proportional to the change in the drain-source current through the following relation: 
\begin{align}
	\label{eq:receptortocurrent}
	N_R(t) = \Delta I_{ds} (t)/Q_{tDNA},
\end{align}
with $Q_{tDNA} = g_m \frac{q_{tDNA}}{C_G}$. Here, $g_m = \partial I_{ds}/ \partial V_g$ is the transconductance of the device, $q_{tDNA}$ is the effective charge of a single tDNA molecule, and $C_{G,0.01\text{xPBS}}$ is the total gate capacitance in 0.01xPBS buffer. Gate capacitance is obtained as  $C_{G,0.01\text{xPBS}} = 6.58\times10^{-2}$ nF using \eqref{capacitance} with the new Debye length in diluted PBS electrolyte, i.e.,  $\lambda_{D,0.01\text{xPBS}} = 7.75$ nm. The effective charge of a single tDNA is obtained similarly using \eqref{effectivecharge} with $\lambda_{D,0.01\text{xPBS}}$. The transconductance of the device $g_m = \partial I_{ds}/ \partial V_g$ before the communication experiments can be calculated by a linear approximation of the $V_g-I_{ds}$ curve around $V_g = 0$ V bias. Averaging over the transfer curves of the four graphene channels obtained at the last step of functionalisation in 0.01xPBS, shown in Fig. \ref{fig:transfer}, we obtain $g_m \approx -28.0 \pm 2.9$ $\mu$A/V. Similarly, the following transformation is made: 
\begin{align}
	\label{eq:receptortocurrenteq}
	N_{R,eq} = \Delta I_{ds,eq}/Q_{tDNA}.
\end{align}
Here, $\Delta I_{ds,eq} $ is the response of the receiver to 1 $\mu$M tDNA at equilibrium, and obtained from the sensing response given in Fig. \ref{fig:sensogram} as $\Delta I_{ds,eq} = -0.805$ $\mu$A. Note that $N_{R,max}$ can also be calculated as a function of $\Delta I_{ds,eq}$ as follows
\begin{align}
	\label{eq:receptortocurrentmax}
	N_{R,max}&= \left((c_{avg}+K_D)/c_{avg}\right) \times N_{R,eq} \\ \nonumber
	&= \left((c_{avg}+K_D)/c_{avg}\right) \times \Delta I_{ds,eq}/Q_{tDNA},
\end{align}
with $c_{avg}$ being the average tDNA concentration passing over the receiver surface \cite{kuscu2018modeling}. The dissociation constant $K_D$ for pDNA-tDNA pair given in Table \ref{table:kinetic}. Accordingly, \eqref{eq:R2} can be rewritten by substituting \eqref{eq:receptortocurrent}, \eqref{eq:receptortocurrenteq}, and \eqref{eq:receptortocurrentmax} into \eqref{eq:R2} as follows
\begin{align} 
	\label{eq:R3}
	\Delta I_{ds} (t) =& \Delta I_{ds,eq} \left( 1- \frac{\mathcal{W}_0\bigl[\alpha^\ast \exp \left(\alpha^\ast - \beta^\ast (t-t_a)\right)\bigr]}{\alpha^\ast}\right) \Big( \Theta\left[t-t_a\right] - \Theta\left[t-t_d-\epsilon \right] \Big) \\ \nonumber
	&- \gamma^\ast Q \mathcal{W}_0 \Bigg[ - \frac{\Delta I_{ds}(t_d)}{\gamma^\ast Q} \exp \left( \frac{-k^+ \Delta I_{ds}(t_d) /Q- k_T^\ast k^- (t-t_d)}{k^+ \gamma^\ast}\right)\Bigg] \Theta\left[t-t_d-\epsilon \right],
\end{align}
with the transformed parameters given as
\begin{equation}
	\alpha^\ast = \frac{k^+ c_{avg} \Delta I_{ds,eq}/Q}{k^- \Delta I_{ds,eq}/Q + k_T^\ast c_{avg}}, \label{alpha3}
\end{equation}
\begin{equation}
	\beta^\ast = \frac{k^+ c_{avg} + k^-}{1 + \frac{k^-  \Delta I_{ds,eq}/Q}{k_T^\ast c_{avg} }}, \label{beta3}
\end{equation}
\begin{equation}
	\gamma^\ast = \frac{(c_{avg} + K_D)  \Delta I_{ds,eq}/Q }{c_{avg}} + \frac{k_T^\ast}{k^+} \label{gamma3}.
\end{equation}

\begin{figure}[!t]
	\centering
	\includegraphics[width=12.5cm]{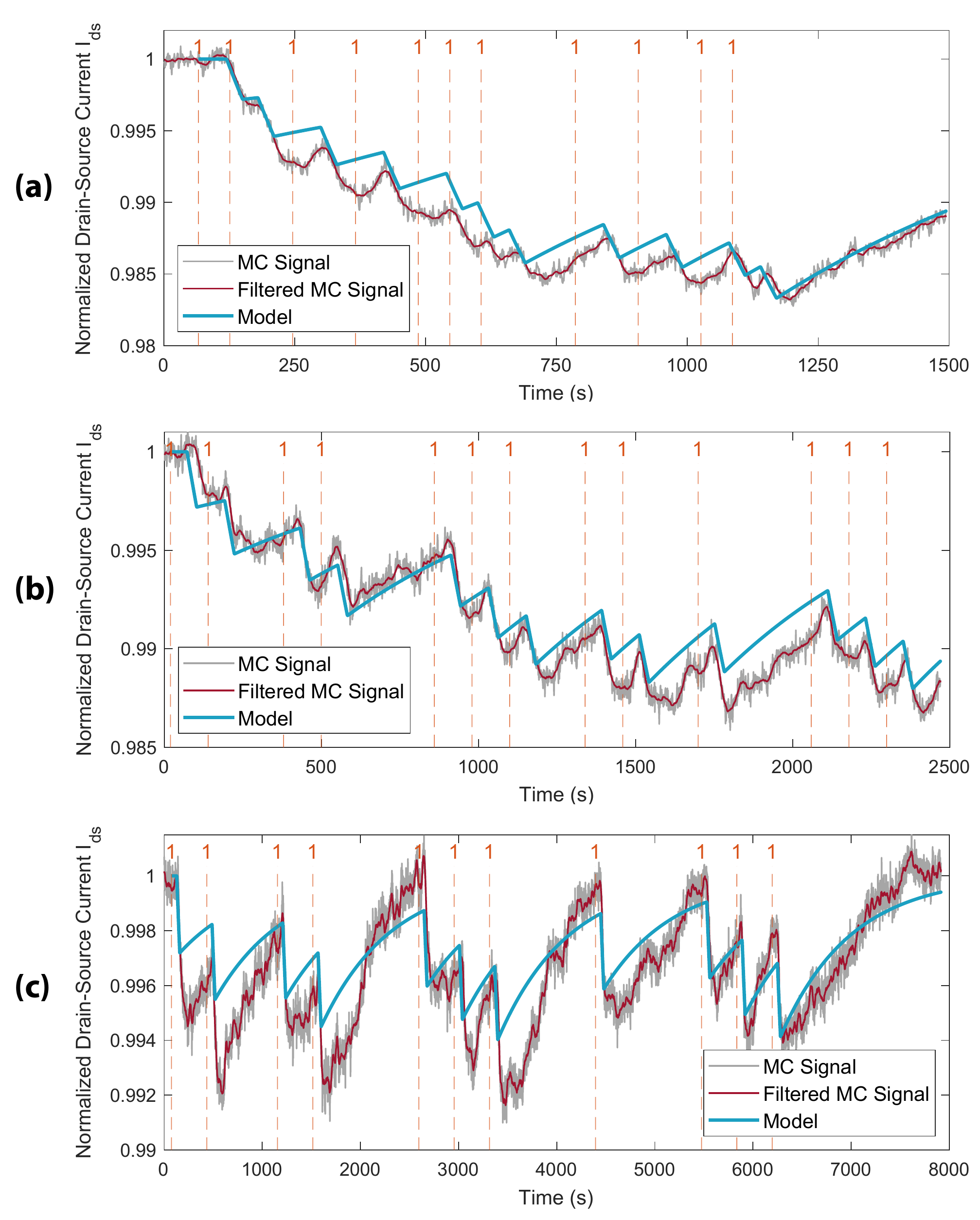}
	\caption{Normalised receiver response for binary data transmission with fixed pulse length $T_p = 30$ s and varying bit intervals: (a) $T_s = 60$ s, (b) $T_s = 120$ s, (c) $T_s = 360$ s. The grey lines denote the received MC signals normalised by the baseline current, and the solid red lines are their low-pass filtered version by a moving average filter of 21-second length in MATLAB. Solid blue lines represent the normalised output of the model given in \eqref{eq:datatransmissionnormalized}. Dashed orange lines indicate the time instants when bit-1 is transmitted, i.e., when the mechanical switch in the tDNA line is opened for 30 seconds.  } 
	\label{fig:comm1}
\end{figure}

The model developed in \cite{kuscu2018modeling} includes many other input parameters concerning the microfluidic channel geometry and molecular transport dynamics. The linear flow velocity is already obtained as $u = 220$ $\mu$m/s by transforming the volumetric flow rate $u_V = 80$ $\mu$l/min. Diffusion coefficient of tDNAs is taken as $D_0 = 100$ $\mu$m$^2$/s, which is in the range of previously reported values for ssDNAs of similar lengths \cite{stellwagen2002determining,nkodo2001diffusion}.

For comparison to the empirical MC signals, the model response is normalised by the mean empirical baseline current ($I_{ds,baseline} = 31.25$ $\mu$A, as plotted in Fig. \ref{fig:sensogram}(a)), i.e., 
\begin{align} 
	\label{eq:datatransmissionnormalized}
	\hat{I}_{ds}(t) = (I_{ds, baseline} + \Delta I_{ds} (t))/I_{ds, baseline}.
\end{align}
The normalised output of the transformed model for pulse lengths $T_p = 30$ s and $T_p = 60$ s is shown in Fig. \ref{fig:pulsefitnew}(c) and Fig. \ref{fig:pulsefitnew}(d), respectively, presented in comparison to the mean of three independent measurements taken for each pulse length. The error bars represent the standard deviation of the measurements. We can observe that the approximate model is highly accurate in calculating the propagation delay and $I_{ds}$ during the association phase. Although its accuracy is lower during the dissociation phase, it well captures the pulse amplitude and the pulse width, which are the two important parameters in evaluating the performance of a communication system. The same model will be applied in the next section for binary data transmission. 

\subsection{Data Transmission}
To reveal the detection performance of the MC receiver, 20-bit-long pseudorandom binary information encoded into the concentration of tDNAs is transmitted through the microfluidic channel and detected by the fabricated MC receiver located at the bottom of the channel. Here bit-1 is represented by a 30-second pulse of $1$ $\mu$M tDNA at the beginning of a bit interval, and bit-0 is represented by no pulse transmission during the entire bit interval, i.e., the buffer line stays connected into the microfluidic channel. The results are provided for bit intervals of 60-, 120-, and 360-second length in Fig. \ref{fig:comm1}. As expected, the slow rate of dissociation of the bound tDNAs from the immobilised pDNAs results in a significant amount of intersymbol interference (ISI), which can potentially complicate the decoding at the receiver. The ISI is more pronounced for shorter bit intervals, i.e., higher transmission rates, such that the baseline could not find enough time to recover. On the other hand, for 360-second-long bit interval (Fig. \ref{fig:comm1}(c)), ISI is significantly lower and $I_{ds}$ is able to return to the baseline.  

Another important observation is that the propagation delay $t_{delay}$ between the transmission time of the bit-1 and the receiver response is $\sim$55 seconds in each test, and the delay remains almost constant during the entire data transmission period. This implies that the application of a constant delay shift in the receiver response can be sufficient for the synchronisation of the receiver with the transmitter during the decoding process.

As the receiver response is away from saturation in the performed tests, we can assume that the response is linear and time-invariant (LTI). Based on this assumption, we can reconstruct the received signal by applying the superposition principle of LTI systems in the approximate model developed in \cite{kuscu2018modeling}. Accordingly, using \eqref{eq:R3}, the received signal can be written as  
\begin{align} 
	\label{eq:datatransmission}
	R(t) = \sum_{i=1}^L s[i] \Delta I_{ds} \left(t-  t_{transmit} - (i-1) T_b \right),
\end{align}
where $L = 20$ is the number of transmitted bits, $t_{transmit}$ is the start time of transmission, $T_b$ is the bit interval, and $s[i] \in \{0,1\}$ is the $i^{\text{th}}$ transmitted bit. Similar to \eqref{eq:datatransmissionnormalized}, the model response is normalised by the mean empirical baseline current ($I_{ds,baseline} = 31.25$ $\mu$A), i.e., 
\begin{align} 
	\label{eq:datatransmissionnormalized2}
	\hat{I}_{ds}(t) = (I_{ds, baseline} + R(t))/I_{ds, baseline}.
\end{align}
The normalised model response is plotted in Fig. \ref{fig:comm1} over the empirical MC signal for varying bit intervals. We can see that the approximate model is very accurate in capturing the transmission delay and the overall trend of $I_{ds}$, although it is not able to exactly reconstruct the original signal. The deviations can largely be attributed to the LTI approximation in writing \eqref{eq:datatransmission}, which neglects the effect of previously transmitted bits on the received signal corresponding to the current bit. 

\begin{figure}[!t]
	\centering
	\includegraphics[width=12.5cm]{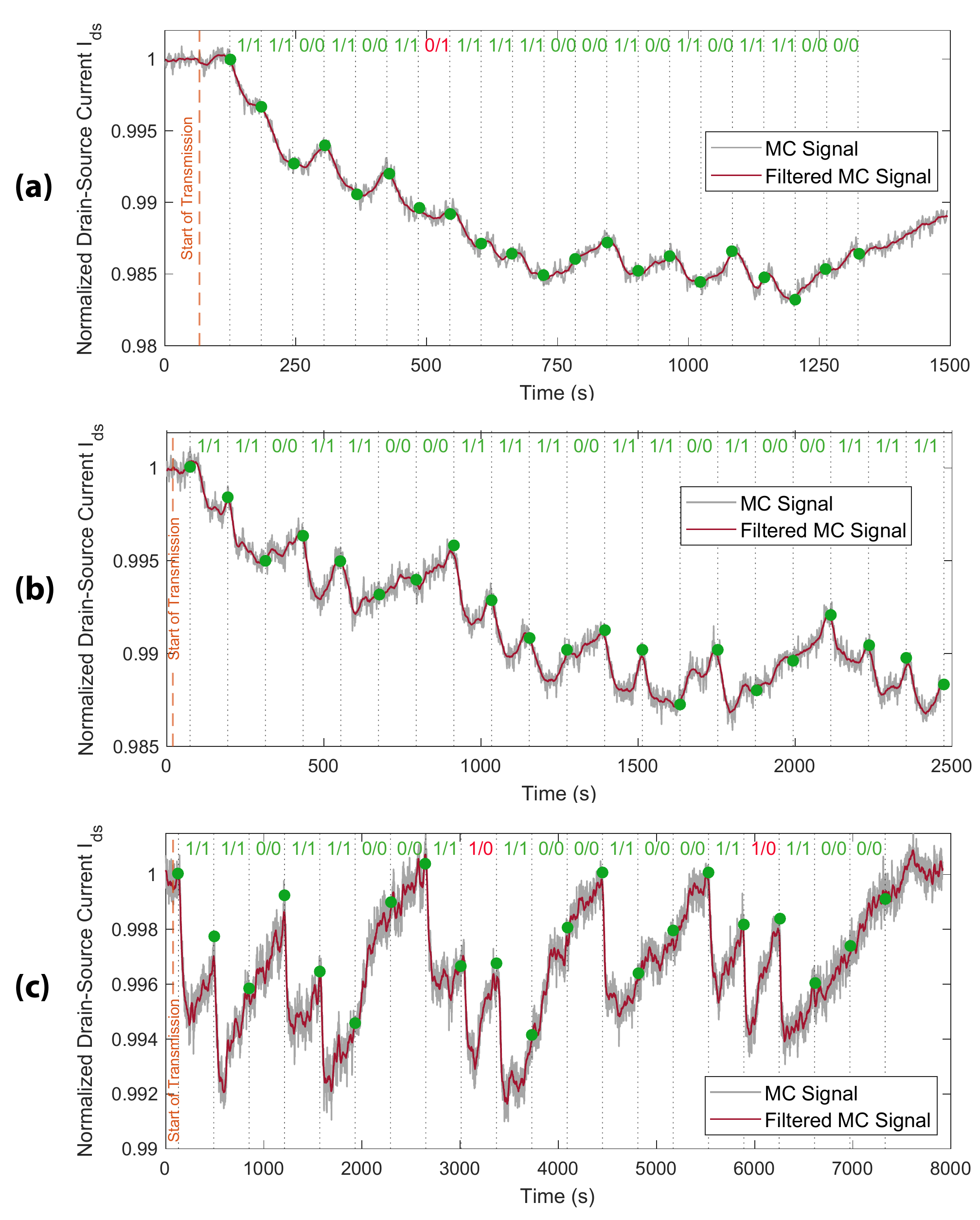}
	\caption{Delay-shifted version of the receiver response for binary data transmission for varying bit intervals: (a) $T_s = 60$ s, (b) $T_s = 120$ s, (c) $T_s = 360$ s. Dashed orange line denotes the start time of the data transmission, and grey dashed lines demarcate the individual bit intervals. Green dots on the received MC signal indicates the sampled current values for difference-based detection method with the decoding rule formulated in \eqref{eq:decoding}. Transmitted/decoded bits are noted above each bit interval, with the red-coloured ones denoting the erroneously decoded bits.   } 
	\label{fig:comm2}
\end{figure}

\begin{figure*}[!t]
	\centering
	\subfigure[]{
		\includegraphics[width=8cm]{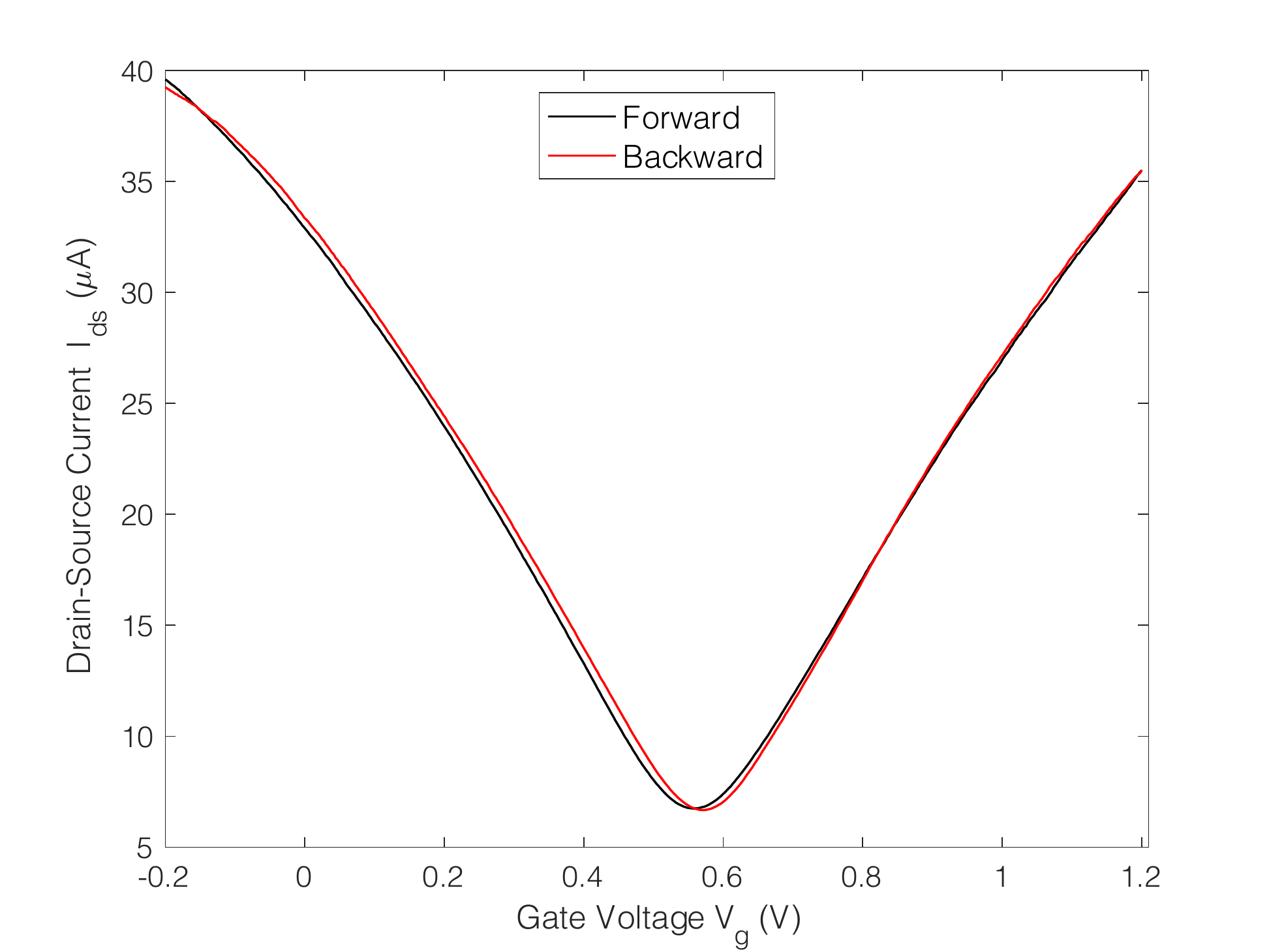}}
	\subfigure[]{
		\includegraphics[width=8cm]{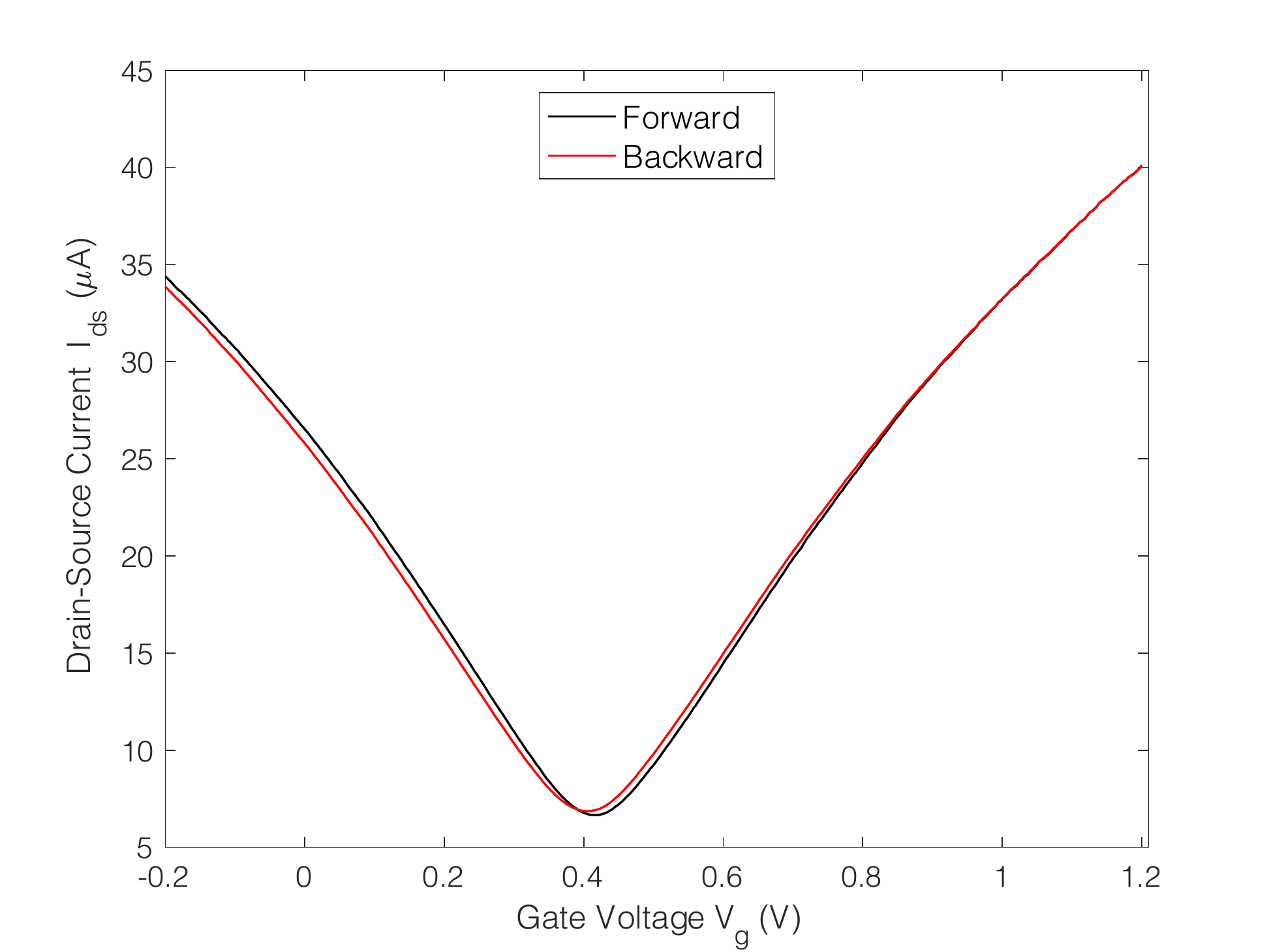}}
	\subfigure[]{
		\includegraphics[width=8cm]{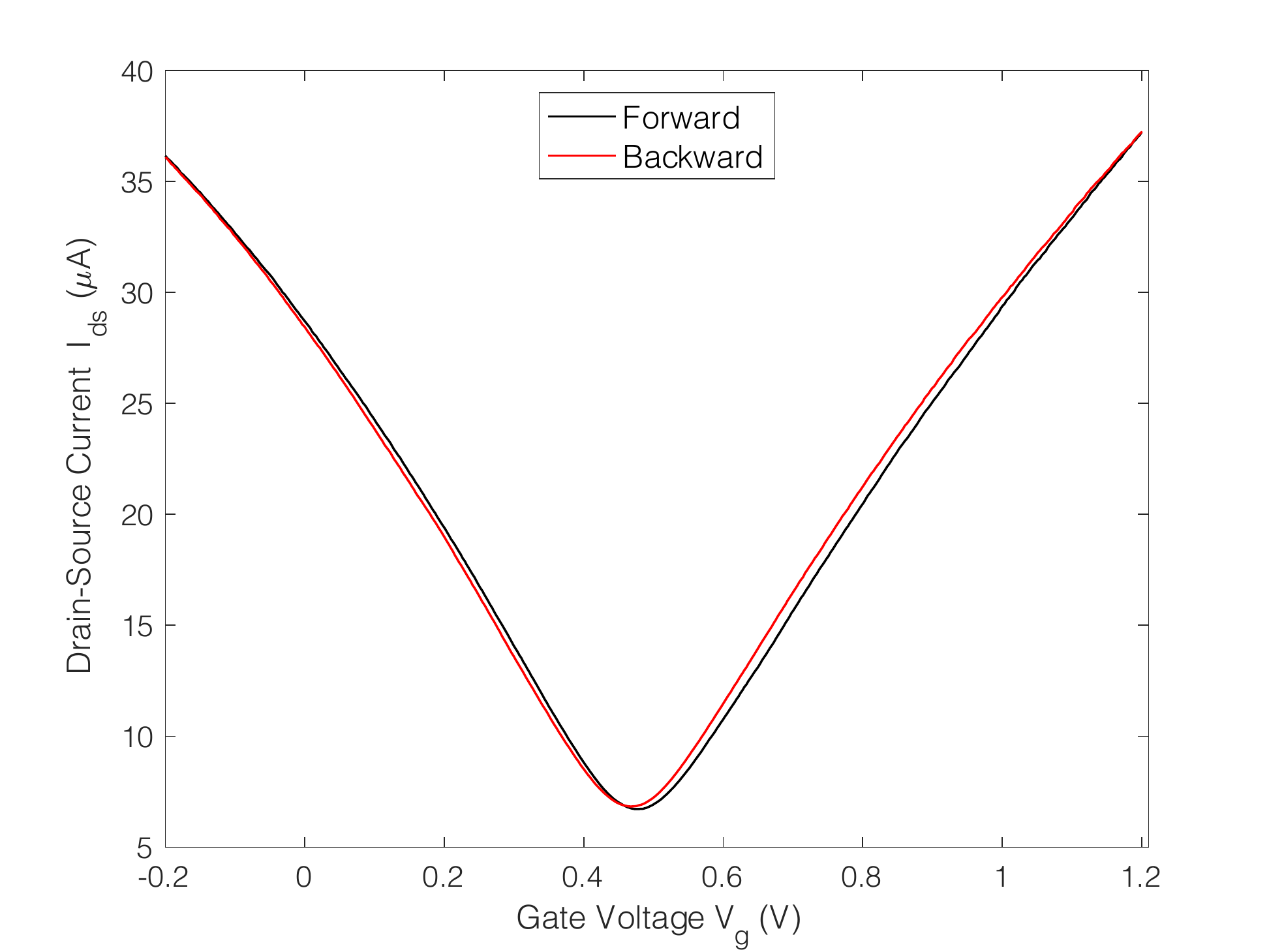}}		
	\subfigure[]{
		\includegraphics[width=8cm]{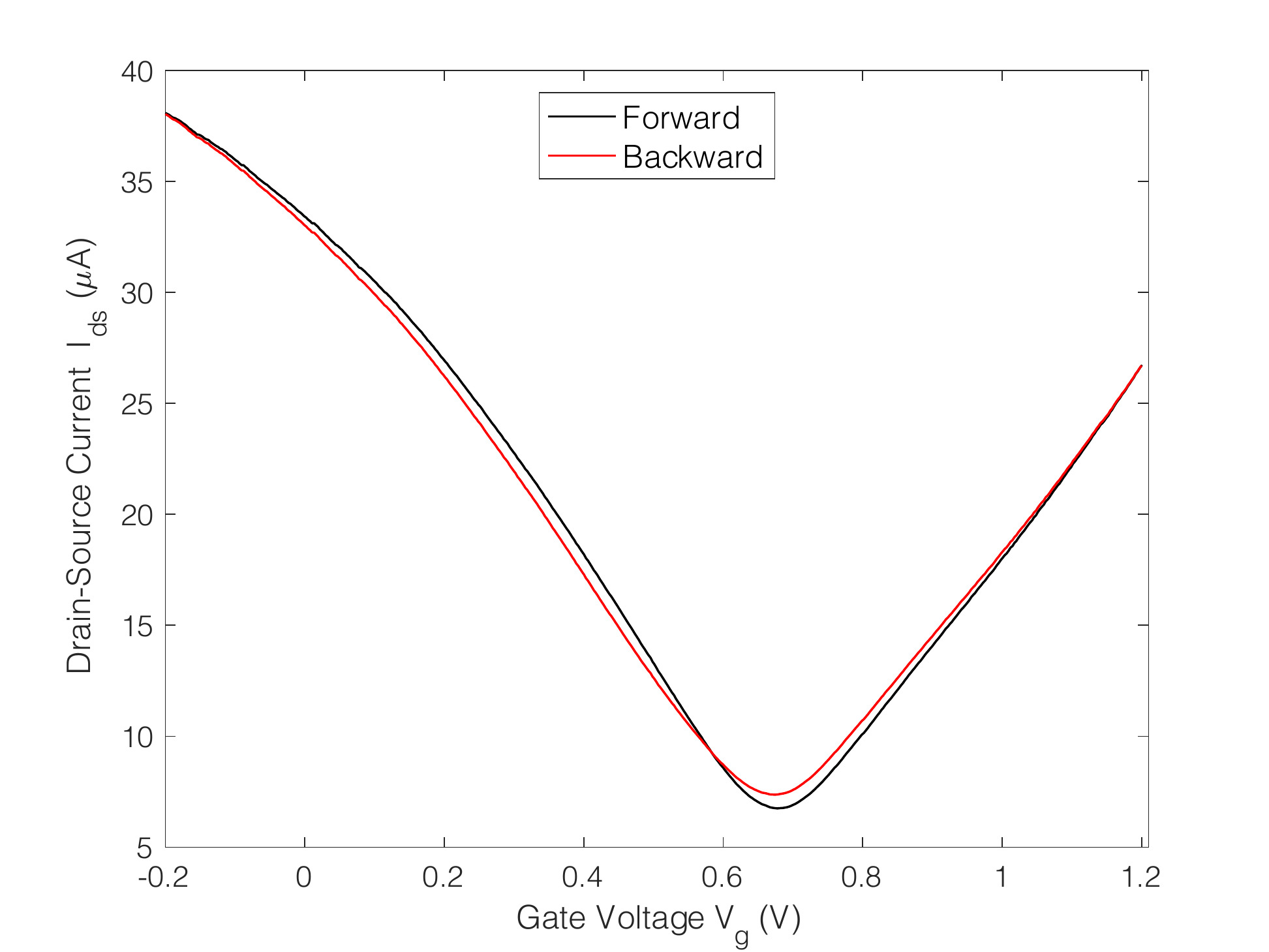}}
	\caption{Hysteresis analysis of the MC receiver: Drain-source current $I_{ds}$ with forward and backward sweep of gate voltage $V_g$ at 140 mV/s sweep rate. (a) Before functionalisation. (b) After functionalisation with PBASE. (c) After immobilization of pDNA. (d) After passivation with ethanolamine.}
	\label{fig:hysteresis}
\end{figure*}

\begin{figure}[h]
	\centering
	\includegraphics[width=11cm]{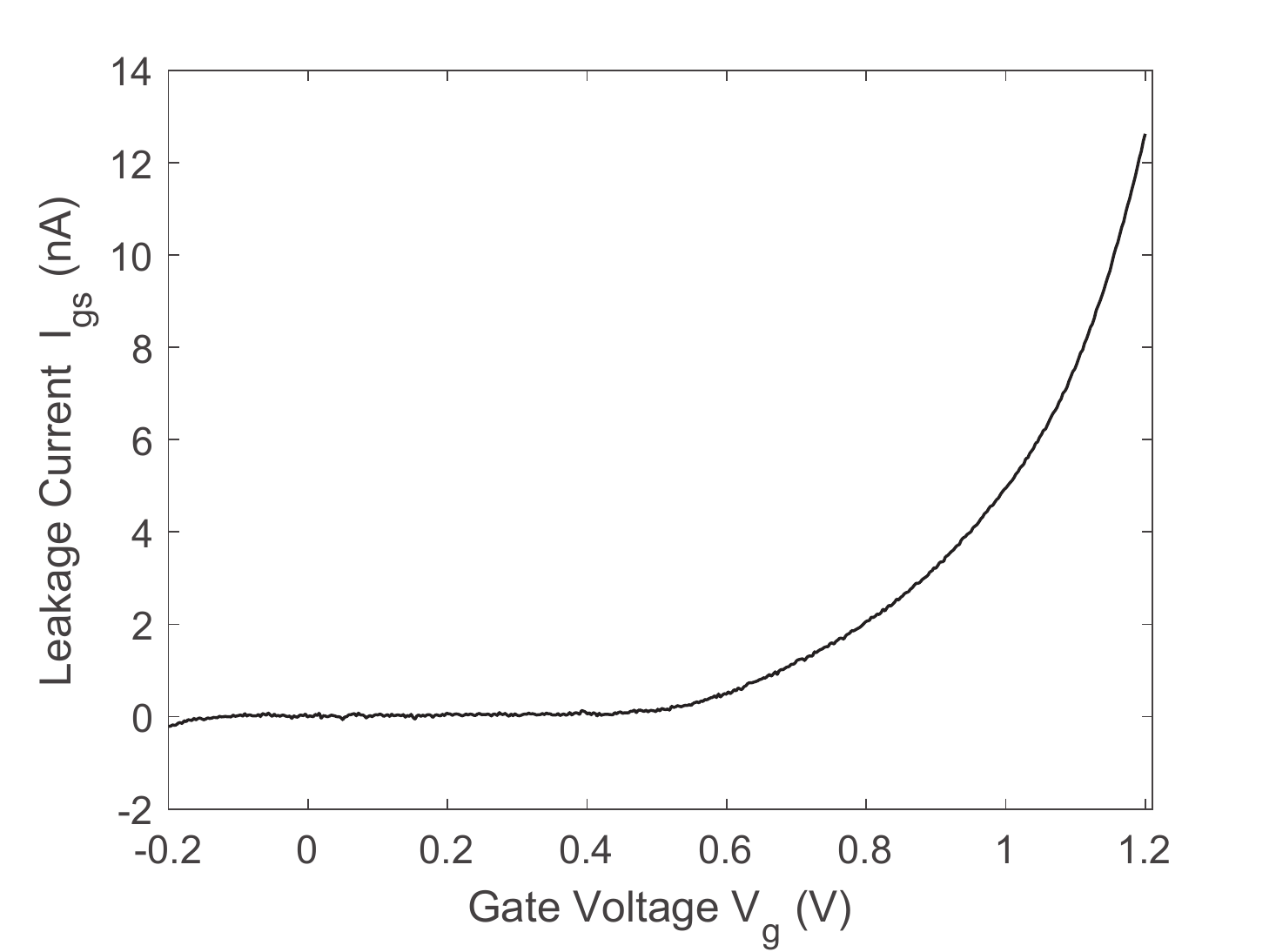}
	\caption[Leakage current analysis of the MC receiver]{Leakage current analysis of the MC receiver: Gate-source current $I_{gs}$ with varying gate voltage $V_g$. }
	\label{fig:leakage}
\end{figure}
Based on the observation of constant delay during data transmission, the delay-shifted version of the bit intervals is indicated with dashed lines in Fig. \ref{fig:comm2} with the transmitted bits written inside each bit interval. As the envisioned MC applications demand low-complexity communication techniques due to the resource and size limitations of the communicating nanodevices, constant threshold detection has been favoured in the literature. In this scheme, the received signal is sampled at a predefined sampling time instant, and its amplitude is compared to a threshold value for deciding between bit-0 and bit-1. However, the high ISI of the MC channel, and the resulting drift of baseline, observed in these experiments, render the constant threshold detection methods ineffective, especially at high communication rates. Therefore, in line with some of the previous work in the MC literature concerning ligand-receptor binding systems, we utilise a simple difference-based detection method, which decodes the information based on the difference in $I_{ds}$ measurements taken at the start and end points of a bit interval. Recall that the receiver is synchronised with the transmitter through the application of a constant delay shift at the receiver side. The sampling time points are demonstrated in Fig. \ref{fig:comm2} with green dots. Accordingly, the decoded bits can be written as follows
\begin{align} 
	\label{eq:decoding}
	\hat{s}[i] = \sum_{i=1}^L s[i] \mathds{1}\bigl[r[i+1]-r[i]<0\bigr],
\end{align}
where $\hat{s}[i]$ is the decoded bit, and $r[i] = I_{ds}(t-t_{transmit}-t_{delay} -(i-1)T_s)$ is the discrete time sample of $I_{ds}$ at decision points, and $\mathds{1}\bigl[.\bigr]$ is the indicator function, which outputs 1 if the inside expression is true. 

Applying the difference-based detection method on the unfiltered MC signal yielded $5\%$ bit error rate (BER). We observed 1 bit error for 60 s bit interval, and 2 bit errors for 360 s bit intervals. The decoding of the unfiltered MC signal of 120 s bit interval yielded no error. The erroneous transmissions are indicated in Fig. \ref{fig:comm2}. On the other hand, the difference-based detection applied on the low-pass filtered MC signal correctly decoded all the transmitted bits.

\section{Conclusion}
This proof-of-concept study reports the very first fabrication and characterisation of a nanoscale graphene bioFET-based MC receiver. The ICT tests of the MC receiver is performed in a custom-designed micro/nanoscale MC system using a PDMS-based microfluidic channel as the controllable propagation medium, and a pressure-regulated flow control system for transmitting binary data encoded into the concentration of target DNA molecules. The time-varying hybridisation of the information-carrying target DNAs with the probe DNAs immobilised on the SLG surface of the MC receiver is selectively transduced into a change in drain-source current over the SLG channel. In light of the experimental sensing results, this transduction mechanism is speculated to be related to both electrostatic gating and direct electron transfer. The ICT performance of the MC receiver is evaluated by binary data transmissions at different bit intervals. The response of the MC receiver is well-fitted with the previously developed approximate analytical model of microfluidic MC. The slow hybridisation kinetics of DNA molecules is revealed to cause significant ISI. A simple difference-based detection method is shown to overcome the ISI to a significant extent, and provide reliable detection performance. The fabricated graphene-based nanoscale MC receiver and the overall microfluidic MC system can be used as an experimental testbed for probing intricate dynamics of MC, and developing novel communication techniques, transceiver architectures, and applications for MC.

\appendices

%\section{Transformation of Microfluidic MC Model???}
%\label{AppendixA}

\section{Supplementary Information}
\label{AppendixB}
\begin{sidewaysfigure}
	\centering
	\includegraphics[width=22cm]{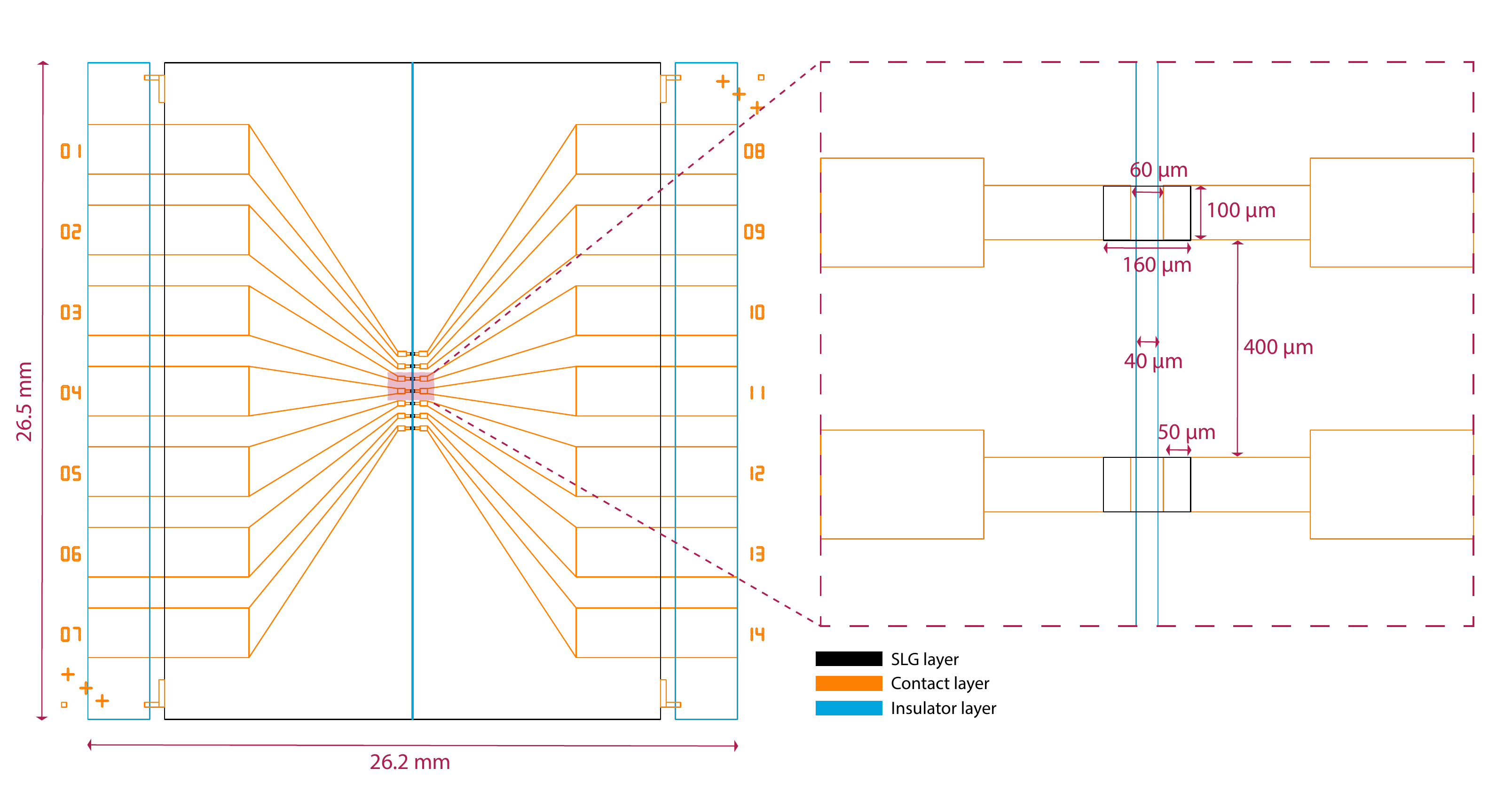}
	\caption{Computer-aided design for optical lithography of MC receiver.}
	\label{fig:litho_layers}
\end{sidewaysfigure}

\begin{sidewaysfigure}
	\centering
	\includegraphics[width=22cm]{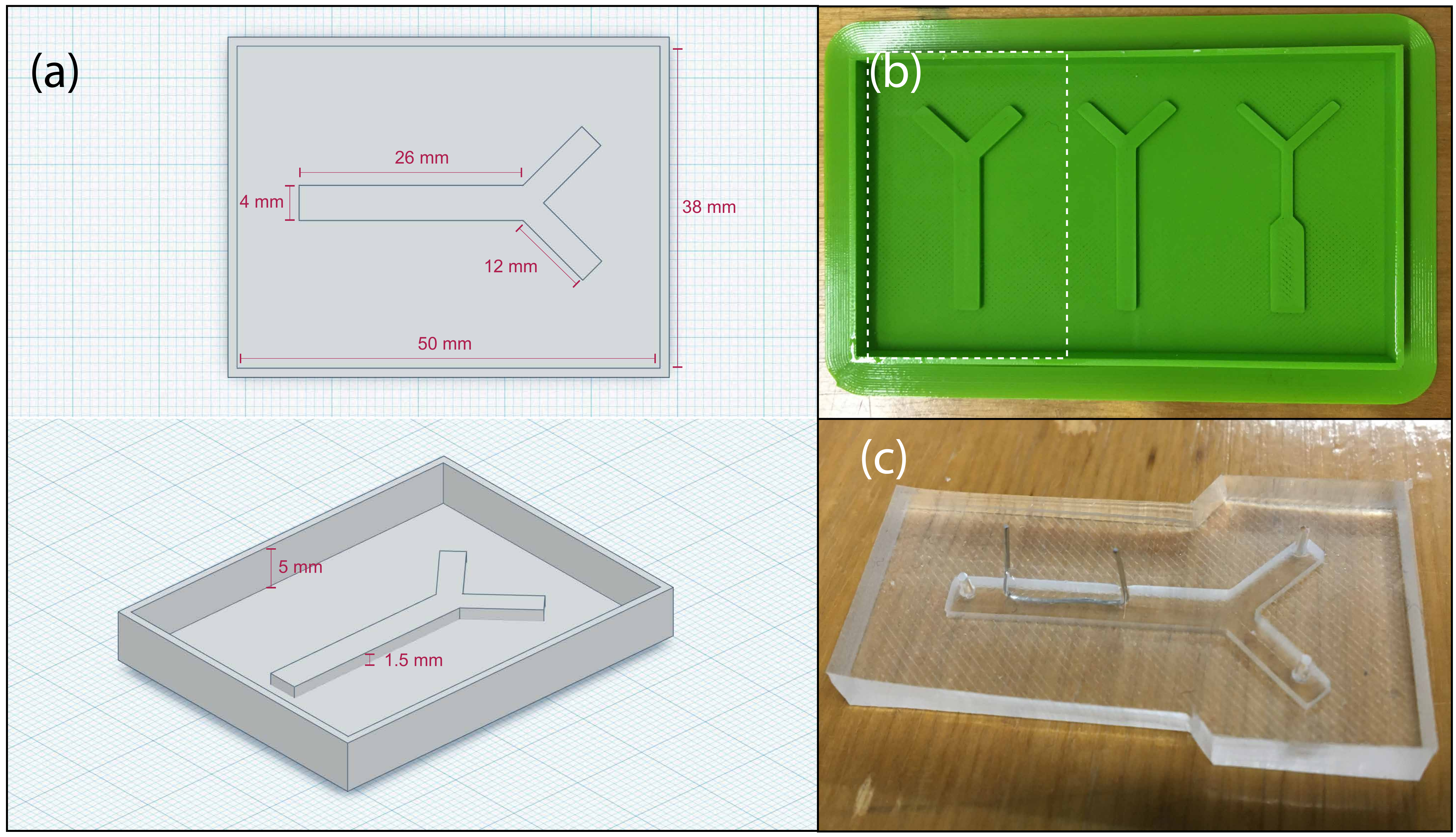}
	\caption{(a) Computer-aided design for 3d printing of PDMS mould. (b) 3d-printed PDMS mould. (c) PDMS microfluidic channel with the Pt wire mounted on top as the gate electrode.}
	\label{fig:PDMS_mould}
\end{sidewaysfigure}

This Appendix provides additional information the hysteresis and leakage current analyses of the MC receiver in Figs. \ref{fig:hysteresis}-\ref{fig:leakage}. Moreover, computer-aided designs (CADs) for the optical lithography of the MC receiver and for the 3d printing of PDMS mould are provided in Figs. \ref{fig:litho_layers}-\ref{fig:PDMS_mould}.

\newpage
\newpage
\newpage
\bibliographystyle{IEEEtran}
\bibliography{references2}

\begin{thebibliography}{10}
\providecommand{\url}[1]{#1}
\csname url@samestyle\endcsname
\providecommand{\newblock}{\relax}
\providecommand{\bibinfo}[2]{#2}
\providecommand{\BIBentrySTDinterwordspacing}{\spaceskip=0pt\relax}
\providecommand{\BIBentryALTinterwordstretchfactor}{4}
\providecommand{\BIBentryALTinterwordspacing}{\spaceskip=\fontdimen2\font plus
\BIBentryALTinterwordstretchfactor\fontdimen3\font minus
  \fontdimen4\font\relax}
\providecommand{\BIBforeignlanguage}[2]{{%
\expandafter\ifx\csname l@#1\endcsname\relax
\typeout{** WARNING: IEEEtran.bst: No hyphenation pattern has been}%
\typeout{** loaded for the language `#1'. Using the pattern for}%
\typeout{** the default language instead.}%
\else
\language=\csname l@#1\endcsname
\fi
#2}}
\providecommand{\BIBdecl}{\relax}
\BIBdecl

\bibitem{akyildiz2010internet}
I.~F. Akyildiz and J.~M. Jornet, ``The internet of nano-things,'' \emph{IEEE
  Wireless Communications}, vol.~17, no.~6, 2010.

\bibitem{akyildiz2015internet}
I.~Akyildiz, M.~Pierobon, S.~Balasubramaniam, and Y.~Koucheryavy, ``The
  {I}nternet of {B}io-{N}ano {T}hings,'' \emph{IEEE Communications Magazine},
  vol.~53, no.~3, pp. 32--40, 2015.

\bibitem{dinc2019internet}
E.~Dinc, M.~Kuscu, B.~A. Bilgin, and O.~B. Akan, ``Internet of everything: A
  unifying framework beyond internet of things,'' in \emph{Harnessing the
  Internet of Everything (IoE) for Accelerated Innovation Opportunities}.\hskip
  1em plus 0.5em minus 0.4em\relax IGI Global, 2019, pp. 1--30.

\bibitem{akan2017fundamentals}
O.~B. Akan, H.~Ramezani, T.~Khan, N.~A. Abbasi, and M.~Kuscu, ``Fundamentals of
  molecular information and communication science,'' \emph{Proceedings of the
  IEEE}, vol. 105, no.~2, pp. 306--318, 2017.

\bibitem{pierobon2011diffusion}
M.~Pierobon, I.~F. Akyildiz \emph{et~al.}, ``Diffusion-based noise analysis for
  molecular communication in nanonetworks,'' \emph{IEEE Transactions on Signal
  Processing}, vol.~59, no.~6, pp. 2532--2547, 2011.

\bibitem{pierobon2011noise}
M.~Pierobon and I.~F. Akyildiz, ``Noise analysis in ligand-binding reception
  for molecular communication in nanonetworks,'' \emph{IEEE Transactions on
  Signal Processing}, vol.~59, no.~9, pp. 4168--4182, 2011.

\bibitem{kuran2011modulation}
M.~S. Kuran, H.~B. Yilmaz, T.~Tugcu, and I.~F. Akyildiz, ``Modulation
  techniques for communication via diffusion in nanonetworks,'' in
  \emph{Communications (ICC), 2011 IEEE International Conference on}.\hskip 1em
  plus 0.5em minus 0.4em\relax IEEE, 2011, pp. 1--5.

\bibitem{kuscu2016modeling}
M.~Kuscu and O.~B. Akan, ``Modeling and analysis of sinw fet-based molecular
  communication receiver,'' \emph{IEEE Transactions on Communications},
  vol.~64, no.~9, pp. 3708--3721, 2016.

\bibitem{bilgin2018dna}
B.~A. Bilgin, E.~Dinc, and O.~B. Akan, ``{DNA}-based molecular
  communications,'' \emph{IEEE Access}, vol.~6, pp. 73\,119--73\,129, 2018.

\bibitem{kuscu2018maximum}
M.~Kuscu and O.~B. Akan, ``Maximum likelihood detection with ligand receptors
  for diffusion-based molecular communications in {I}nternet of {B}io-{N}ano
  {T}hings,'' \emph{IEEE Transactions on Nanobioscience}, vol.~17, no.~1, pp.
  44--54, 2018.

\bibitem{kuscu2019channel}
------, ``Channel sensing in molecular communications with single type of
  ligand receptors,'' \emph{IEEE Transactions on Communications}, vol.~67,
  no.~10, pp. 6868--6884, 2019.

\bibitem{kuscu2019transmitter}
M.~Kuscu, E.~Dinc, B.~A. Bilgin, H.~Ramezani, and O.~B. Akan, ``Transmitter and
  receiver architectures for molecular communications: A survey on physical
  design with modulation, coding, and detection techniques,'' \emph{Proceedings
  of the IEEE}, vol. 107, no.~7, pp. 1302--1341, 2019.

\bibitem{jamali2019channel}
V.~Jamali, A.~Ahmadzadeh, W.~Wicke, A.~Noel, and R.~Schober, ``Channel modeling
  for diffusive molecular communication—a tutorial review,''
  \emph{Proceedings of the IEEE}, vol. 107, no.~7, pp. 1256--1301, 2019.

\bibitem{farsad2013tabletop}
N.~Farsad, W.~Guo, and A.~W. Eckford, ``Tabletop molecular communication: Text
  messages through chemical signals,'' \emph{PloS One}, vol.~8, no.~12, p.
  e82935, 2013.

\bibitem{kim2015universal}
N.-R. Kim, N.~Farsad, C.-B. Chae, and A.~W. Eckford, ``A universal channel
  model for molecular communication systems with metal-oxide detectors,'' in
  \emph{Communications (ICC), 2015 IEEE International Conference on}.\hskip 1em
  plus 0.5em minus 0.4em\relax IEEE, 2015, pp. 1054--1059.

\bibitem{koo2016molecular}
B.-H. Koo, C.~Lee, H.~B. Yilmaz, N.~Farsad, A.~Eckford, and C.-B. Chae,
  ``Molecular mimo: From theory to prototype,'' \emph{IEEE Journal on Selected
  Areas in Communications}, vol.~34, no.~3, pp. 600--614, 2016.

\bibitem{farsad2017novel}
N.~Farsad, D.~Pan, and A.~Goldsmith, ``A novel experimental platform for
  in-vessel multi-chemical molecular communications,'' in \emph{GLOBECOM
  2017-2017 IEEE Global Communications Conference}.\hskip 1em plus 0.5em minus
  0.4em\relax IEEE, 2017, pp. 1--6.

\bibitem{unterweger2018experimental}
H.~Unterweger, J.~Kirchner, W.~Wicke, A.~Ahmadzadeh, D.~Ahmed, V.~Jamali,
  C.~Alexiou, G.~Fischer, and R.~Schober, ``Experimental molecular
  communication testbed based on magnetic nanoparticles in duct flow,'' in
  \emph{2018 IEEE 19th International Workshop on Signal Processing Advances in
  Wireless Communications (SPAWC)}.\hskip 1em plus 0.5em minus 0.4em\relax
  IEEE, 2018, pp. 1--5.

\bibitem{wicke2018magnetic}
W.~Wicke, A.~Ahmadzadeh, V.~Jamali, H.~Unterweger, C.~Alexiou, and R.~Schober,
  ``Magnetic nanoparticle-based molecular communication in microfluidic
  environments,'' \emph{IEEE Transactions on Nanobioscience}, vol.~18, no.~2,
  pp. 156--169, 2019.

\bibitem{ferrari2015science}
A.~C. Ferrari, F.~Bonaccorso, V.~Fal'Ko, K.~S. Novoselov, S.~Roche,
  P.~B{\o}ggild, S.~Borini, F.~H. Koppens, V.~Palermo, N.~Pugno \emph{et~al.},
  ``Science and technology roadmap for graphene, related two-dimensional
  crystals, and hybrid systems,'' \emph{Nanoscale}, vol.~7, no.~11, pp.
  4598--4810, 2015.

\bibitem{huang2010nanoelectronic}
Y.~Huang, X.~Dong, Y.~Shi, C.~M. Li, L.-J. Li, and P.~Chen, ``Nanoelectronic
  biosensors based on {CVD} grown graphene,'' \emph{Nanoscale}, vol.~2, no.~8,
  pp. 1485--1488, 2010.

\bibitem{zhan2014graphene}
B.~Zhan, C.~Li, J.~Yang, G.~Jenkins, W.~Huang, and X.~Dong, ``Graphene
  field-effect transistor and its application for electronic sensing,''
  \emph{Small}, vol.~10, no.~20, pp. 4042--4065, 2014.

\bibitem{kwak2012flexible}
Y.~H. Kwak, D.~S. Choi, Y.~N. Kim, H.~Kim, D.~H. Yoon, S.-S. Ahn, J.-W. Yang,
  W.~S. Yang, and S.~Seo, ``Flexible glucose sensor using cvd-grown
  graphene-based field effect transistor,'' \emph{Biosensors and
  Bioelectronics}, vol.~37, no.~1, pp. 82--87, 2012.

\bibitem{ohno2009electrolyte}
Y.~Ohno, K.~Maehashi, Y.~Yamashiro, and K.~Matsumoto, ``Electrolyte-gated
  graphene field-effect transistors for detecting ph and protein adsorption,''
  \emph{Nano Letters}, vol.~9, no.~9, pp. 3318--3322, 2009.

\bibitem{xu2017real}
S.~Xu, J.~Zhan, B.~Man, S.~Jiang, W.~Yue, S.~Gao, C.~Guo, H.~Liu, Z.~Li,
  J.~Wang \emph{et~al.}, ``Real-time reliable determination of binding kinetics
  of {DNA} hybridization using a multi-channel graphene biosensor,''
  \emph{Nature Communications}, vol.~8, p. 14902, 2017.

\bibitem{campos2019attomolar}
R.~Campos, J.~Borme, J.~R. Guerreiro, G.~Machado~Jr, M.~F. Cerqueira, D.~Y.
  Petrovykh, and P.~Alpuim, ``Attomolar label-free detection of dna
  hybridization with electrolyte-gated graphene field-effect transistors,''
  \emph{ACS Sensors}, vol.~4, no.~2, pp. 286--293, 2019.

\bibitem{kuscu2016physical}
M.~Kuscu and O.~B. Akan, ``On the physical design of molecular communication
  receiver based on nanoscale biosensors,'' \emph{IEEE Sensors Journal},
  vol.~16, no.~8, pp. 2228--2243, 2016.

\bibitem{green2015interactions}
N.~S. Green and M.~L. Norton, ``Interactions of dna with graphene and sensing
  applications of graphene field-effect transistor devices: A review,''
  \emph{Analytica Chimica Acta}, vol. 853, pp. 127--142, 2015.

\bibitem{malak2012molecular}
D.~Malak and O.~B. Akan, ``Molecular communication nanonetworks inside human
  body,'' \emph{Nano Communication Networks}, vol.~3, no.~1, pp. 19--35, 2012.

\bibitem{chahibi2014molecular}
Y.~Chahibi and I.~F. Akyildiz, ``Molecular communication noise and capacity
  analysis for particulate drug delivery systems,'' \emph{IEEE Transactions on
  Communications}, vol.~62, no.~11, pp. 3891--3903, 2014.

\bibitem{deng2015modeling}
Y.~Deng, A.~Noel, M.~Elkashlan, A.~Nallanathan, and K.~C. Cheung, ``Modeling
  and simulation of molecular communication systems with a reversible
  adsorption receiver,'' \emph{IEEE Transactions on Molecular, Biological and
  Multi-Scale Communications}, vol.~1, no.~4, pp. 347--362, 2015.

\bibitem{li2015low}
B.~Li, M.~Sun, S.~Wang, W.~Guo, and C.~Zhao, ``Low-complexity noncoherent
  signal detection for nanoscale molecular communications,'' \emph{IEEE
  Transactions on Nanobioscience}, vol.~15, no.~1, pp. 3--10, 2015.

\bibitem{chueh2007leakage}
B.-h. Chueh, D.~Huh, C.~R. Kyrtsos, T.~Houssin, N.~Futai, and S.~Takayama,
  ``Leakage-free bonding of porous membranes into layered microfluidic array
  systems,'' \emph{Analytical Chemistry}, vol.~79, no.~9, pp. 3504--3508, 2007.

\bibitem{bicen2013system}
A.~O. Bicen and I.~F. Akyildiz, ``System-theoretic analysis and least-squares
  design of microfluidic channels for flow-induced molecular communication,''
  \emph{IEEE Transactions on Signal Processing}, vol.~61, no.~20, pp.
  5000--5013, 2013.

\bibitem{yue2017electricity}
W.~Yue, C.~Tang, C.~Wang, C.~Bai, S.~Liu, X.~Xie, H.~Hua, Z.~Zhang, and D.~Li,
  ``An electricity-fluorescence double-checking biosensor based on graphene for
  detection of binding kinetics of dna hybridization,'' \emph{RSC Advances},
  vol.~7, no.~70, pp. 44\,559--44\,567, 2017.

\bibitem{hwang2016highly}
M.~T. Hwang, P.~B. Landon, J.~Lee, D.~Choi, A.~H. Mo, G.~Glinsky, and R.~Lal,
  ``Highly specific snp detection using 2d graphene electronics and dna strand
  displacement,'' \emph{Proceedings of the National Academy of Sciences}, vol.
  113, no.~26, pp. 7088--7093, 2016.

\bibitem{wu2017doping}
G.~Wu, X.~Tang, M.~Meyyappan, and K.~W.~C. Lai, ``Doping effects of surface
  functionalization on graphene with aromatic molecule and organic solvents,''
  \emph{Applied Surface Science}, vol. 425, pp. 713--721, 2017.

\bibitem{kabelavc2012influence}
M.~Kabel{\'a}{\v{c}}, O.~Kroutil, M.~P{\v{r}}edota, F.~Lanka{\v{s}}, and
  M.~{\v{S}}{\'\i}p, ``Influence of a charged graphene surface on the
  orientation and conformation of covalently attached oligonucleotides: a
  molecular dynamics study,'' \emph{Physical Chemistry Chemical Physics},
  vol.~14, no.~12, pp. 4217--4229, 2012.

\bibitem{ghosh2018selective}
S.~Ghosh, N.~I. Khan, J.~G. Tsavalas, and E.~Song, ``selective detection of
  lysozyme biomarker utilizing large area chemical vapor deposition-grown
  graphene-based field-effect transistor,'' \emph{Frontiers in Bioengineering
  and Biotechnology}, vol.~6, p.~29, 2018.

\bibitem{chen2013label}
T.-Y. Chen, P.~T.~K. Loan, C.-L. Hsu, Y.-H. Lee, J.~T.-W. Wang, K.-H. Wei,
  C.-T. Lin, and L.-J. Li, ``Label-free detection of dna hybridization using
  transistors based on cvd grown graphene,'' \emph{Biosensors and
  Bioelectronics}, vol.~41, pp. 103--109, 2013.

\bibitem{zheng2013electrical}
Y.~Zheng, J.~Nguyen, C.~Wang, and Y.~Sun, ``Electrical measurement of red blood
  cell deformability on a microfluidic device,'' \emph{Lab on a Chip}, vol.~13,
  no.~16, pp. 3275--3283, 2013.

\bibitem{goldsmith2019digital}
B.~R. Goldsmith, L.~Locascio, Y.~Gao, M.~Lerner, A.~Walker, J.~Lerner, J.~Kyaw,
  A.~Shue, S.~Afsahi, D.~Pan \emph{et~al.}, ``Digital biosensing by
  foundry-fabricated graphene sensors,'' \emph{Scientific Reports}, vol.~9,
  no.~1, pp. 1--10, 2019.

\bibitem{meric2008current}
I.~Meric, M.~Y. Han, A.~F. Young, B.~Ozyilmaz, P.~Kim, and K.~L. Shepard,
  ``Current saturation in zero-bandgap, top-gated graphene field-effect
  transistors,'' \emph{Nature Nanotechnology}, vol.~3, no.~11, p. 654, 2008.

\bibitem{manoharan2017simplified}
A.~K. Manoharan, S.~Chinnathambi, R.~Jayavel, and N.~Hanagata, ``Simplified
  detection of the hybridized {DNA} using a graphene field effect transistor,''
  \emph{Science and Technology of Advanced Materials}, vol.~18, no.~1, pp.
  43--50, 2017.

\bibitem{dong2010electrical}
X.~Dong, Y.~Shi, W.~Huang, P.~Chen, and L.-J. Li, ``Electrical detection of dna
  hybridization with single-base specificity using transistors based on
  cvd-grown graphene sheets,'' \emph{Advanced Materials}, vol.~22, no.~14, pp.
  1649--1653, 2010.

\bibitem{xu2014electrophoretic}
G.~Xu, J.~Abbott, L.~Qin, K.~Y. Yeung, Y.~Song, H.~Yoon, J.~Kong, and D.~Ham,
  ``Electrophoretic and field-effect graphene for all-electrical dna array
  technology,'' \emph{Nature Communications}, vol.~5, p. 4866, 2014.

\bibitem{star2006label}
A.~Star, E.~Tu, J.~Niemann, J.-C.~P. Gabriel, C.~S. Joiner, and C.~Valcke,
  ``Label-free detection of dna hybridization using carbon nanotube network
  field-effect transistors,'' \emph{Proceedings of the National Academy of
  Sciences}, vol. 103, no.~4, pp. 921--926, 2006.

\bibitem{van2008see}
J.~van Mameren, E.~J. Peterman, and G.~J. Wuite, ``See me, feel me: methods to
  concurrently visualize and manipulate single dna molecules and associated
  proteins,'' \emph{Nucleic Acids Research}, vol.~36, no.~13, pp. 4381--4389,
  2008.

\bibitem{kaiser2010conformations}
W.~Kaiser and U.~Rant, ``Conformations of end-tethered dna molecules on gold
  surfaces: influences of applied electric potential, electrolyte screening,
  and temperature,'' \emph{Journal of the American Chemical Society}, vol. 132,
  no.~23, pp. 7935--7945, 2010.

\bibitem{rant2004dynamic}
U.~Rant, K.~Arinaga, S.~Fujita, N.~Yokoyama, G.~Abstreiter, and M.~Tornow,
  ``Dynamic electrical switching of dna layers on a metal surface,'' \emph{Nano
  Letters}, vol.~4, no.~12, pp. 2441--2445, 2004.

\bibitem{kuscu2018modeling}
M.~Kuscu and O.~B. Akan, ``Modeling convection-diffusion-reaction systems for
  microfluidic molecular communications with surface-based receivers in
  {I}nternet of {B}io-{N}ano {T}hings,'' \emph{PloS One}, vol.~13, no.~2, p.
  e0192202, 2018.

\bibitem{stellwagen2002determining}
E.~Stellwagen and N.~C. Stellwagen, ``Determining the electrophoretic mobility
  and translational diffusion coefficients of dna molecules in free solution,''
  \emph{Electrophoresis}, vol.~23, no.~16, pp. 2794--2803, 2002.

\bibitem{nkodo2001diffusion}
A.~E. Nkodo, J.~M. Garnier, B.~Tinland, H.~Ren, C.~Desruisseaux, L.~C.
  McCormick, G.~Drouin, and G.~W. Slater, ``Diffusion coefficient of dna
  molecules during free solution electrophoresis,'' \emph{Electrophoresis},
  vol.~22, no.~12, pp. 2424--2432, 2001.

\end{thebibliography}

\end{document}